\RequirePackage{silence}
 \documentclass[10pt,aps,prx,twocolumn,superscriptaddress,nofootinbib,longbibliography,floatfix]{revtex4-2}

\usepackage[english]{babel} 

\usepackage[T1]{fontenc}

\usepackage{amsmath}
\usepackage{amsfonts}
\usepackage{bm}         
\usepackage{physics}    

\usepackage{booktabs}   
\usepackage{multirow}   
\usepackage{dcolumn}    

\usepackage{graphicx}

\graphicspath{{figures/}} 
\usepackage{hyperref}
\hypersetup{
    colorlinks=true,
    linkcolor=blue,
    citecolor=blue,
    urlcolor=blue
}

\begin{document}

\title{Phase controlled spectral topology, dynamic stability and sensitivity in Non-Hermitian Cavity Magnonics}

\author{Sachin Singh Bargahi}
\email{sachinsinghbargahi.phy23@itbhu.ac.in}
\author{Rajeev Singh}
\email{Corresponding author, E-mail: rajeevs.phy@iitbhu.ac.in}

\author{Biswanath Bhoi}
\email{Corresponding author, E-mail: biswanath.phy@itbhu.ac.in}

\affiliation{Nano-Magnetism and Quantum Technology Lab,
 Department of Physics, Indian Institute of Technology (Banaras Hindu University) Varanasi, Varanasi - 221005, India.}

\date{\today}

\begin{abstract}
We theoretically investigate a non-Hermitian cavity-magnon platform in which coherent photon-magnon interactions and reservoir-mediated dissipative coupling interfere through a single externally tunable phase. We show that this interference phase provides a universal control parameter that continuously rotates the effective coupling between Hermitian and anti-Hermitian regimes, enabling dynamic transitions between level repulsion and level attraction without modifying intrinsic system parameters. The resulting phase-controlled non-Hermitian topology gives rise to exceptional points, linewidth engineering, and zero-damping conditions. Owing to the propagation-direction dependence of the dissipative interaction, the system further exhibits strong nonreciprocal transport and phase-tunable isolation arising from asymmetric hybridization of the cavity and magnon modes.

Beyond its spectral and transport properties, we establish a direct connection between non-Hermitian spectral topology and nonequilibrium population dynamics. The interference phase governs the stability of the hybrid modes, driving transitions between stable relaxation, critical slowing down near exceptional points, oscillatory energy exchange, and exponentially amplified dynamics. We further demonstrate that the same phase-controlled exceptional topology can be exploited for enhanced sensing, where the eigenvalue response exhibits the characteristic square-root scaling associated with exceptional-point physics. Our results provide a unified framework linking spectral topology, directional transport, dynamical stability, and sensing functionality through reservoir-engineered interference in cavity magnonic systems.

 \vspace{1em}

\begin{description}
\item[Usage]

\item[Structure]
 \texttt{description}  
\end{description}
\end{abstract}

 
\maketitle

\section{Introduction}

Light-matter interaction plays a central role in quantum information science and technology because coherent exchange of quantum states is required for practical information transfer\cite{Quantum_technologies_with_hybrid_systems,feynman1982,shor1997,divincenzo2001,Wallquist_2009,Kimble_2008,Xiang2013}. Hybrid quantum systems aim to merge the complementary advantages of disparate quantum platforms into a unified architecture, enabling storage, processing, and transmission of quantum information\cite{Clerk:2020aro_HQS_cQED}. This integration enables functionalities that surpass the limitations of individual subsystems. Photons, for instance, offer exceptional coherence properties and are ideal carriers of quantum information. On the other hand, magnons quanta of spin waves in magnetically ordered systems provide localized\cite{LachanceQuirion2017}, collective excitations that are highly tunable via external magnetic fields \cite{Huebl2013,Tabuchi2014,Lachance-Quirion_2019}. A typical cavity magnonic system consists of a ferromagnetic material, such as yttrium-iron-garnet (YIG), placed inside a microwave cavity where it interacts with the confined electromagnetic field. When these elements are coupled coherently\cite{Control_of_the_Magnon_Photon_Coupling,Huebl2013,Tabuchi2014,Zhang2014,Bai2015}, they form hybrid modes that allow the manipulation, conversion, and storage of quantum information in novel ways, and when coupled dissipatively, they allow inducing non-Hermitian behavior that provides control over loss, phase, and energy flow, giving rise to phenomena such as level attraction and exceptional points and thus extending the capabilities of hybrid quantum systems\cite{PhysRevB.99.134426Abnormal_anticrossing_effect_in_photon-magnon_coupling, Magnon-Photon_Level_Attraction_repulsion}. Cavity magnonics, which combines magnons and microwave cavity photons, is a rapidly developing field within its broader landscape\cite{BHOI20191}. While coherent photon–magnon coupling has enabled numerous advances in hybrid quantum systems, recent studies have revealed that dissipation can also play a constructive role by inducing effective interactions through a common reservoir. Such reservoir-mediated interactions lead naturally to non-Hermitian physics \cite{Bender1998,Bender_2007,ElGanainy2018,Ashida2020NonHermitianP,Bergholtz2021}, where gain, loss, and interference become integral components of system dynamics . In contrast to conventional Hermitian systems, non-Hermitian cavity magnonic systems can exhibit level attraction, linewidth bifurcation, exceptional points (EPs), and nonreciprocal transport \cite{Heiss2012,Chen2017,Miri2019}. Understanding how these phenomena can be controlled in a unified manner remains an important challenge for both fundamental studies and practical applications.

In practical implementations, it is often desirable to access both coherent and dissipative coupling regimes within a single system, as each regime supports different functionalities, and the dynamical stability and steady-state behavior of driven photon-magnon systems remain insufficiently understood. In many previously reported cavity magnonic systems, transitions between coherent and dissipative coupling are achieved by modifying the cavity geometry, changing the loading conditions, or engineering the spatial overlap between cavity and magnon modes. Such approaches often require structural modifications and therefore provide limited tunability during device operation. A more versatile strategy is to manipulate the relative phase between coherent and dissipative interaction pathways. Since coherent and dissipative couplings represent orthogonal interaction channels, a controllable phase difference allows continuous rotation of the effective interaction in the complex plane. Consequently, the coupling can be tuned from predominantly Hermitian to predominantly anti-Hermitian behavior without altering any intrinsic material or device parameters.
Furthermore, in particular, understanding how interactions influence the transition between stable saturation and dynamical instability, as well as the redistribution of energy between hybridized modes, remains a challenge. Traditionally, this transition is achieved by modifying system parameters such as coupling geometry, external loading, or intrinsic loss rates. However, altering these physical components during device operation is not always feasible. To address this limitation, we consider a framework based on complex coupling, where the interaction strength acquires both real and imaginary components. This approach enables continuous tuning between coherent and dissipative regimes through external control parameters, such as phase or drive conditions, without requiring any structural modifications. While recent studies have demonstrated the emergence of non-Hermitian effects under such conditions, a systematic exploration of phase-controlled tunability and its impact on system dynamics remains limited \cite{Anomalous_coherent_and_dissipative_coupling,in_device_level_attraction_repulsion, LA_LR_anti_resonance, Combination_of_coherent_dissipative_coupling}. The phase associated with complex coupling provides a unified mechanism to tune the system behavior consistently across a variety of magnetic materials.

In this work, we investigate a driven cavity--magnon system incorporating complex, phase-dependent coupling and examine its influence on both transient and steady-state dynamics. Starting from an effective Hamiltonian, we derive the corresponding equations of motion and obtain analytical expressions for the steady-state amplitudes and populations of the photon and magnon modes. Our analysis reveals that the system exhibits qualitatively distinct behaviors governed by the coupling phase, including stable saturation in the absence of gain and exponential amplification when stability conditions are not satisfied. Furthermore, we show that the coupling phase serves as an efficient control parameter for tuning the steady-state populations of the hybrid modes. Through a detailed investigation of the time evolution, we demonstrate that stable regimes correspond to bounded population dynamics, whereas unstable regimes lead to unbounded growth. We also examine the role of detuning in shaping the steady-state response, highlighting the intricate interplay between phase, coupling strength, and frequency mismatch. Unlike previous studies that investigated spectral topology, exceptional points, nonreciprocal transport, or population dynamics separately, the present work demonstrates that all these phenomena emerge from a common phase-controlled complex coupling mechanism. We show that a single interference phase simultaneously governs the nature of photon–magnon interaction, the formation of exceptional points, directional transport characteristics, and dynamical stability. This establishes a unified framework for controlling non-Hermitian cavity magnonic systems through reservoir-engineered interference.

This paper is organized as follows. Section II introduces the theoretical model and derives the effective non-Hermitian Hamiltonian. Section III examines the phase-dependent transition between coherent and dissipative interaction regimes. Section IV investigates directional transport and isolation arising from propagation-dependent interference. Section V establishes the connection between spectral topology and dynamical stability. Section VI analyzes photon and magnon population dynamics near exceptional points. Finally, Section VII summarizes the main conclusions and discusses future opportunities for phase-engineered cavity magnonic devices. Overall, our results indicate that the interplay between intrinsic dissipation and phase-dependent coupling determines the system dynamics, enabling controlled transitions between level repulsion and level attraction. These findings identify the coupling phase as a key resource for manipulating energy flow, stability, and mode hybridization in cavity magnonic systems and, more generally, in open quantum systems exhibiting non-Hermitian interactions.

\section{Theoretical Model}
The system under consideration consists of a microwave cavity mode (photon mode) coupled to a magnon mode in a ferromagnetic material (e.g., YIG). The cavity is driven by an external input field $p_{\mathrm{in}}(t)$ through an input port characterized by a coupling rate $\kappa_{\mathrm{in}}$. The output field $p_{\mathrm{out}}(t)$ is measured from the cavity. The intrinsic cavity loss and magnon damping are denoted by $\beta$ and $\alpha$, respectively. The coupling coefficient of cavity and magnon modes from the external bath are $\beta$ and $\kappa$, respectively.  The coherent and dissipative interaction between the photon and magnon modes is described by the coupling strength g and $\Gamma e^{i(\Theta+\Phi)}$, respectively.
\begin{figure*}
		\centering		\includegraphics[width=0.75\linewidth]{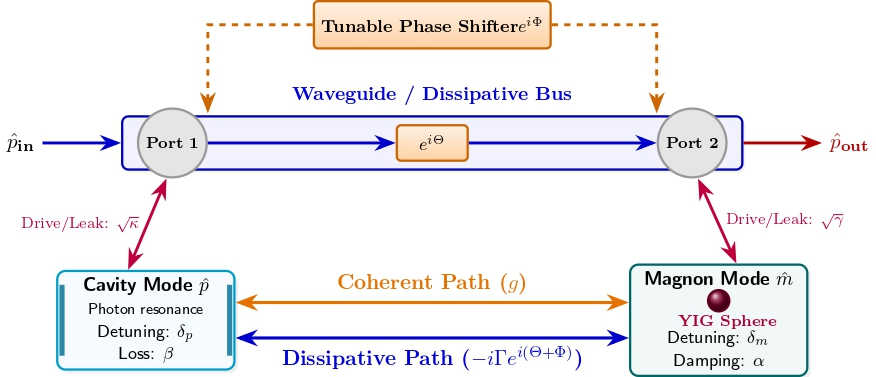}
		\caption{
\textbf{Schematic of the non-Hermitian cavity-magnon platform.}
A common waveguide simultaneously drives the cavity photon mode ($\hat{p}$) and YIG magnon mode ($\hat{m}$), generating coherent coupling $g$ and reservoir-mediated dissipative coupling $-i\Gamma e^{i(\Theta+\Phi)}$. The propagation phase $\Theta$ and external phase $\Phi$ provide tunable control of the complex coupling and the resulting output field $\hat{p}_{\mathrm{out}}$.
}
		\label{fig:theoretical-schematicfigure}
	\end{figure*}
Under the rotating wave approximation (RWA) derived in Appendix (\ref{app:RWA_hamiltonian}), the coherent Hamiltonian is expressed as follows:

\begin{align}
\hat{H} = \delta_p \hat{p}^\dagger \hat{p} + \delta_m \hat{m}^\dagger \hat{m}
+ g \left( \hat{p}^\dagger \hat{m} +\hat{m}^\dagger \hat{p}   \right)
\end{align}
where $\delta_p = \omega_p - \omega$ and $\delta_m = \omega_m - \omega$ denote the detuning of the cavity and magnon frequencies relative to the input drive frequency $\omega$, respectively. The parameter $g$ represents the coherent photon-magnon coupling strength. The last term describes the coherent driving field applied to the cavity.

To account for the losses in the cavity and magnon subsystems, we incorporate damping into the Hamiltonian formalism by extending the detuning terms to complex values. This leads to a non-Hermitian effective Hamiltonian, whose eigenvalues are generally complex, thereby encoding both the resonance frequencies and the decay rates of the hybridized modes.

The modified detuning for both modes is given  by $\tilde{\delta}_p = \delta_p - i\beta$ and $\tilde{\delta}_m = \delta_m - i\alpha$.

In any realistic quantum system, interaction with the surrounding environment leads to dissipation and decoherence, so to account for this non-Hermitian system, we use Lindblad master formalism\cite{Lindblad_master_equation}, which modifies coupling as $  {g} - i\Gamma e^{i(\Theta+\Phi)}$. Details are provided in  Appendix(\ref{app:dissipation and damping in H RWA}) 
\begin{equation}
\begin{aligned}
\hat H =&\,
\tilde{\delta}_p \hat p^\dagger \hat p
+\tilde{\delta}_m \hat m^\dagger \hat m \\
&+\left(g-i\Gamma e^{i(\Theta+\Phi)}\right)
\left(\hat p^\dagger \hat m
+\hat m^\dagger \hat p\right) \\
&+ i\sqrt{\kappa}\,p_{\rm in}
(\hat p^\dagger-\hat p)
+i\sqrt{\gamma}\,p_{\rm in}
(\hat m^\dagger-\hat m).
\end{aligned}
\label{Final_Hamiltonian}
\end{equation}
 The effective interaction term contains two physically distinct contributions. The real coupling parameter g originates from a coherent direct exchange of energy between cavity photons and magnons, whereas the imaginary contribution $ \Gamma e^{i(\Theta+\Phi)}$ arises from reservoir-mediated dissipation. The propagation phase $ \Theta$ and externally tunable phase $ \Phi$ determine the interference between these two interaction pathways. Consequently, the effective coupling can be continuously rotated in the complex plane, providing direct access to both coherent and dissipative interaction regimes.

The dynamics of the annihilation operator of both photon and magnon modes can be analyzed using the master-equation formalism. This involves solving the coupled dynamical equations obtained from the Heisenberg–Langevin formalism derived in Equations (\ref{Appendix_dynamics_of_photon}) and (\ref{Appendix_dynamics_of_magnon}), incorporating damping and coherent interaction. Using the commutation relations of the creation and annihilation operators that apply for both modes,
$[\hat{p}_i, \hat{p}_j^\dagger] = \delta_{ij}, \quad
[\hat{p}_i, \hat{p}_j] = [\hat{p}_i^\dagger, \hat{p}_j^\dagger] = 0,$

The dynamics of the photon and magnon modes from the Hamiltonian described in the \eqref{Final_Hamiltonian} by using the Von Neumann-Heisenberg equation 
\begin{subequations}
\begin{align}
\dot{\hat{p}}
&=
(-\beta-i\delta_p)\hat{p}
-i\left(g-i\Gamma e^{i(\Theta+\Phi)}\right)\hat{m}
+\sqrt{\kappa}\,p_{\mathrm{in}},
\\
\dot{\hat{m}}
&=
(-\alpha-i\delta_m)\hat{m}
-i\left(g-i\Gamma e^{i(\Theta+\Phi)}\right)\hat{p}
+\sqrt{\gamma}\,p_{\mathrm{in}}.
\end{align}
\label{Mode_dynamics}
\end{subequations}
can be written in Equations (\ref{Mode_dynamics})a and (\ref{Mode_dynamics})b in the matrix form for analysis of the resonance frequency and linewidths 
 \begin{align*}
\frac{d}{dt}
\begin{pmatrix}
\hat{p} \\
\hat{m}
\end{pmatrix}
=
- i \, H_{\mathrm{eff}}
\begin{pmatrix}
\hat{p} \\
\hat{m}
\label{Dyanamics_matrix_equation}
\end{pmatrix}
+
\begin{pmatrix}
\sqrt{\kappa} \\
\sqrt{\gamma}
\end{pmatrix}
\hat{p}_{\mathrm{in}}
\end{align*}

\textbf{Eigenfrequencies and damping  Characteristic}
For the analysis of the behavior of the intrinsic mode and damping, we consider the coupled cavity–magnon Hamiltonian excluding external drive. Hamiltonian $H_{eff}$:

\begin{equation}
H_{eff}=
\begin{pmatrix}
 \delta_p-i\beta  & {g} - i\Gamma e^{i(\Theta+\Phi)} \\
{g} - i\Gamma e^{i(\Theta+\Phi)} & \delta_m-i\alpha
\end{pmatrix}.
\label{Eigenvalue_matrix}
\end{equation}

The eigenvalues of this matrix are
\begin{equation}
\begin{split}
  \lambda_{\pm} = {} & \frac{1}{2} \left[ \delta_p + \delta_m - i (\beta + \alpha) \right] \\
  & \pm \frac{1}{2} \sqrt{ \left[\delta_p - \delta_m - i (\beta - \alpha) \right]^2 + 4\left(g - i\Gamma e^{i(\Theta+\Phi)}\right)^2 }
\end{split}
\label{Eigen_value_equation}
\end{equation}

The complex eigenvalues contain complete information regarding the hybridized photon–magnon modes. Their real parts determine the resonance frequencies of the polariton branches, whereas their imaginary parts describe the effective damping rates. Therefore, the spectral evolution of the eigenvalues directly reveals the competition between coherent energy exchange and dissipative mode hybridization. As the interference phase varies, the system can transition continuously between level repulsion and level attraction, eventually reaching exceptional points where eigenvalues and eigenvectors simultaneously coalesce. These complex eigenvalues can be written as $\lambda_{\pm} = \omega_{\pm} - i \Gamma_{\pm}$, where
$\omega_{\pm}$ are the effective hybrid resonance frequencies and 2$\Gamma_{\pm}$ are the associated
line widths (or decay rates) of the upper and lower polariton branches. The real part of the square root
term determines the degree of mode splitting, while the imaginary part governs the broadening due to dissipation.

In the strong coupling regime, where $g > \frac{|\beta - \alpha|}{4}$, the eigenfrequencies exhibit a clear
\textit{normal-mode splitting} (level repulsion) at $\phi=0$, and level attraction, $\phi=\pi$, which is observable in the transmission spectrum. Conversely,
in the weak coupling regime, the square root becomes predominantly imaginary, leading to overlapping peaks
with no resolvable splitting.

\section{PHASE-CONTROLLED TRANSITION BETWEEN COHERENT AND DISSIPATIVE INTERACTIONS}

The central concept of this work is that the interference phase controls the balance between coherent and dissipative coupling channels. Since the coherent interaction is associated with the real part of the effective coupling and the dissipative interaction is associated with its imaginary part, the phase effectively determines the relative weight of Hermitian and anti-Hermitian interactions. This mechanism allows the interaction landscape to be continuously tuned without changing the physical structure of the system. The complex effective coupling governs the nature of the interaction in the system,
\begin{equation}
G = g - i\Gamma e^{i(\Theta+\Phi)} ,
\end{equation}
which incorporates both coherent and dissipative contributions. By separating real and imaginary parts, the coupling can be written as
\begin{align}
\mathrm{Re}(G_) &= g + \Gamma \sin(\Theta+\Phi), \\
\mathrm{Im}(G) &= -\Gamma \cos(\Theta+\Phi).
\end{align}
The real part describes coherent (Hermitian) energy exchange between the modes, whereas the imaginary part captures dissipative (non-Hermitian) coupling mediated by the environment.

To quantify the relative strength of these two mechanisms, we define the dimensionless ratio.
\begin{equation}
R(\Phi) = \log_{10}\left(\frac{|\mathrm{Im}(G)|}{|\mathrm{Re}(G)|}\right).
\end{equation}
When $R < 0$, the interaction is dominated by coherent coupling, while $R > 0$ corresponds to dissipative coupling dominance. The condition $R = 0$ marks the crossover point where both contributions are equal. The coherent-dissipative crossover has profound consequences for the system spectrum. In the coherent-dominated regime, the hybridized modes exhibit level repulsion and normal-mode splitting. As the dissipative contribution increases, the frequency splitting gradually decreases while linewidth splitting increases. At the crossover boundary, the system approaches an exceptional point, beyond which dissipative interactions dominate, and level attraction emerges. Therefore, the phase-controlled interaction ratio serves as a direct indicator of the underlying spectral topology.
\begin{figure}
    \centering
    \includegraphics[width=1\linewidth]{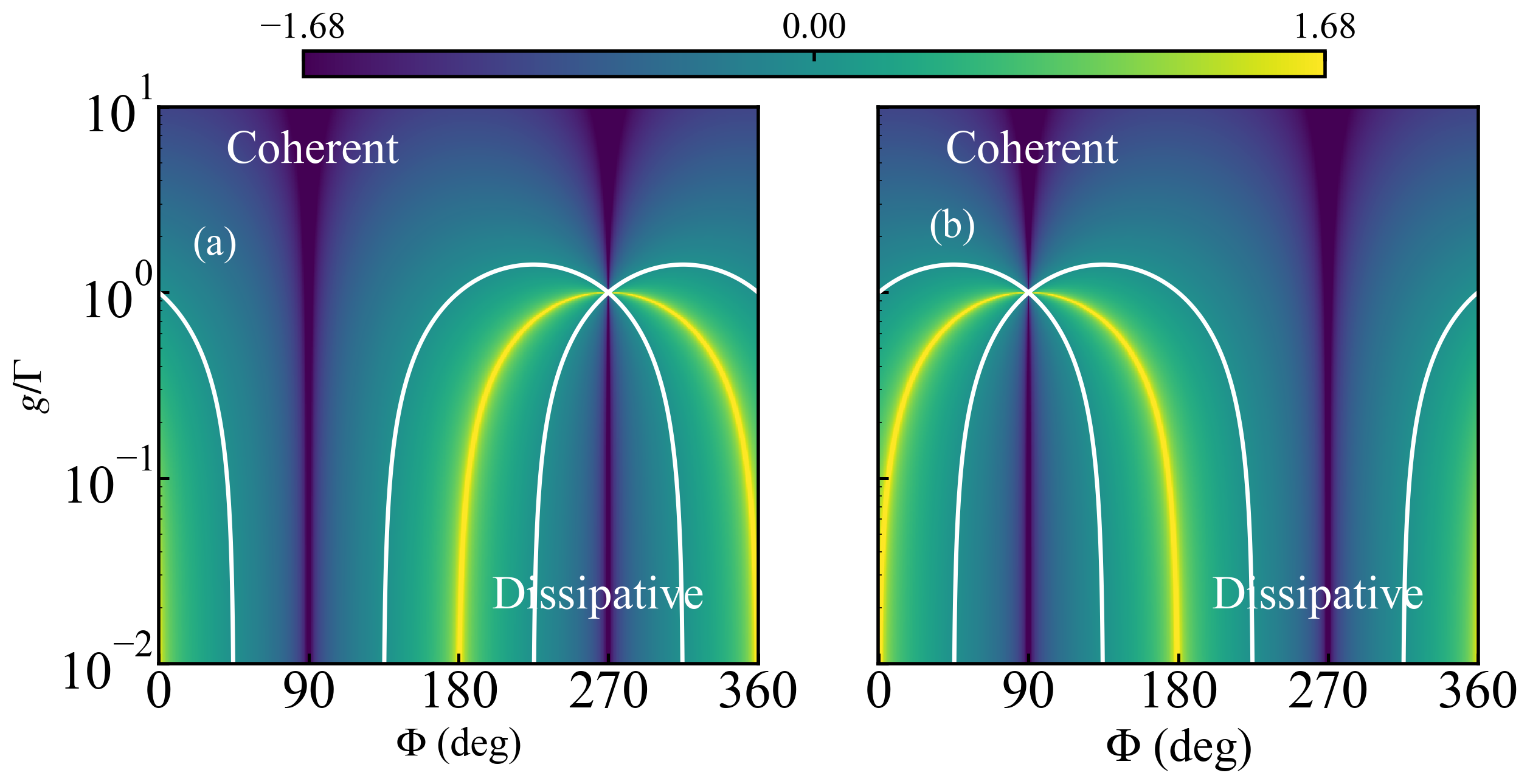}
    \caption{
Phase diagram of the effective cavity--magnon interaction in the $(\Phi,g/\Gamma)$ parameter space. The color scale shows
$R(\Phi)=\log_{10}|\mathrm{Im}(G)/\mathrm{Re}(G)|$, which measures the ratio between dissipative and coherent coupling strengths. Regions with $R>0$ are dissipative dominated, whereas regions with $R<0$ are coherent dominated. The white contour corresponds to $R=0$, where coherent and dissipative interactions contribute equally. The contour branches intersect at $(\Phi,g/\Gamma)=(90^\circ,1)$ and $(270^\circ,1)$, identifying the exceptional-point conditions of the non-Hermitian cavity--magnon system. These points separate the coherent- and dissipative-dominated interaction regimes, providing a natural boundary for the emergence of exceptional-point dynamics.
}
    \label{fig:coherent_dissipative_coupling}
\end{figure}

As the phase $\Phi$ is varied, the system undergoes a continuous and periodic transition between coherent and dissipative regimes. This demonstrates that $\Phi$ acts as an effective external control parameter, enabling dynamic tuning of the interaction without modifying intrinsic system parameters. In particular, the ability to shift the balance point ($R=0$) with propagation direction highlights the role of phase engineering in achieving nonreciprocal behavior.

\section{Phase-Controlled Isolation and Directional Transport}

The origin of nonreciprocity in the system lies in the direction-dependent phase $\Theta$ entering the effective coupling
\begin{equation}
G = g - i\Gamma e^{i(\Theta+\Phi)}.
\end{equation}

The origin of nonreciprocal transport can be understood as an interference phenomenon between coherent and dissipative coupling pathways. Because the dissipative interaction acquires opposite phases for forward and backward propagation, the effective coupling becomes direction-dependent. Consequently, constructive interference occurring in one propagation direction transforms into destructive interference in the opposite direction. This asymmetry modifies the degree of mode hybridization and gives rise to unequal transmission amplitudes. For forward and backward propagation, the phase takes the values $\Theta=0$ and $\Theta=\pi$, respectively. This fundamental asymmetry leads to different effective couplings in the two directions:
\begin{equation}
G_f = g + \Gamma \sin\Phi - i\Gamma \cos\Phi, 
\label{farward_effective_coupling}
\end{equation}
\begin{equation}
G_b = g - \Gamma \sin\Phi + i\Gamma \cos\Phi.
\label{backward_effective_coupling}
\end{equation}

For forward propagation ($\Theta= 0$), the coherent and dissipative contributions combine according to Eq. \eqref{farward_effective_coupling}, whereas for backward propagation ($\Theta=\pi$), the dissipative contribution changes sign as shown in Eq. \eqref{backward_effective_coupling}. As a result, the effective interaction strength experienced by the hybrid modes differs in the two directions. This directional asymmetry constitutes the fundamental mechanism responsible for the observed nonreciprocal transmission. As a result, the balance between coherent and dissipative interactions becomes direction dependent, giving rise to nonreciprocal transport ($S_{21} \neq S_{12}$). Importantly, while the intrinsic system parameters remain unchanged, the external phase $\Phi$ acts as a control knob that continuously tunes the system between different interaction regimes.
\subsection{Isolation and Directional Transport}

Exceptional points play a crucial role in enhancing nonreciprocal transport. Near an EP, small variations in the effective coupling produce large changes in the hybrid mode structure. Since the EP locations differ for forward and backward propagation, one direction can approach mode coalescence while the other remains strongly hybridized. This asymmetry amplifies the difference between ($S_{21}$) and ($S_{12}$), resulting in enhanced isolation. The degree of nonreciprocity is quantified by the isolation ratio.
\begin{equation}
\mathrm{Isolation} = 20 \log_{10} \left|\frac{S_{21}}{S_{12}}\right|.
\end{equation}
Where the transmission amplitude is given by this formula derived in Appendix \ref{app:Transmission Coefficient}.
\begin{align}
S_{Trs}(\Theta) = \frac{\hat{p}_{\mathrm{out}}}{\hat{p}_{\mathrm{in}}}=
1
+
\frac{\kappa A_m + \gamma A_p + 2\sqrt{\kappa\gamma}\,G_{\mathrm{eff}}}
{A_pA_m-G_{\mathrm{eff}}^2} .
\end{align}

Since the direction-dependent effective coupling directly governs the transmission amplitudes, the isolation strongly depends on the phase $\Phi$. In particular, when the coherent and dissipative couplings become comparable $(g \sim \Gamma)$, the interference between the two interaction channels becomes highly asymmetric for opposite propagation directions. This asymmetric interference can strongly suppress transmission in one direction while simultaneously enhancing it in the reverse direction, resulting in large isolation ratios and near-unidirectional transport.

The phase-dependent complex coupling also significantly modifies the non-Hermitian eigenspectrum of the system. Exceptional points (EPs) emerge when both the eigenvalues and eigenvectors simultaneously coalesce. For the effective two-mode Hamiltonian, the EP condition is determined from the vanishing of the discriminant.
\begin{equation}
D=a_r^2-4b=0 .
\label{eq:EPcondition}
\end{equation}

Near the EPs, the system exhibits strong spectral sensitivity accompanied by linewidth redistribution and enhanced mode asymmetry. Because of the propagation-dependent phase, the effective hybridization becomes markedly different for the forward and backward directions in the vicinity of the EPs. Consequently, one direction experiences enhanced mode coupling and transmission, whereas the opposite direction remains weakly hybridized, producing strongly asymmetric transport and enhanced isolation.
\begin{figure}
    \centering
\includegraphics[width=1\linewidth]{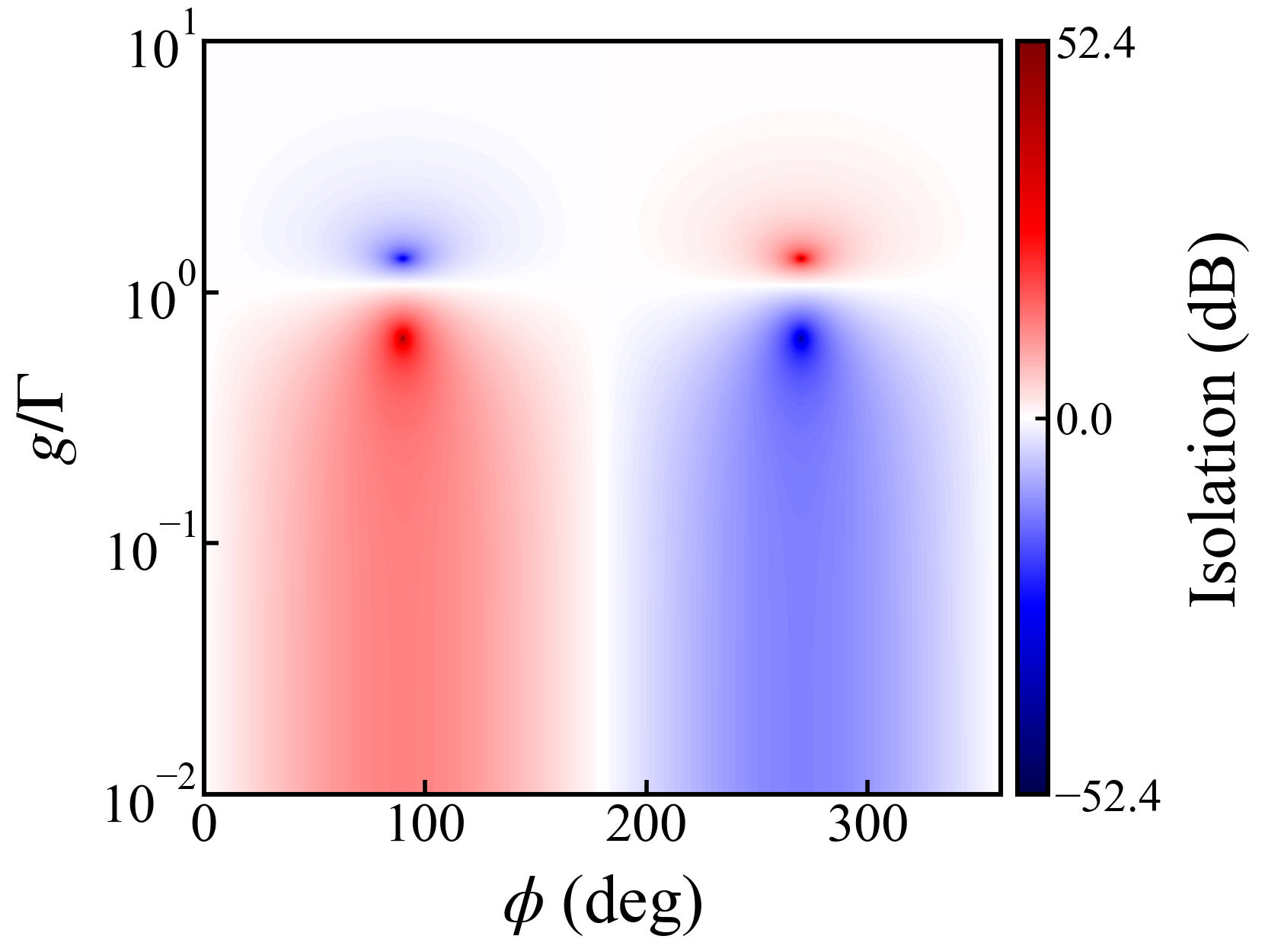}
    \caption{
Isolation ratio as a function of the coupling phase $\Phi$ and normalized coupling strength $g/\Gamma$. The colormap represents the isolation.
$
\mathrm{Isolation}
=
20\log_{10}
\left|
\frac{S_{21}}{S_{12}}
\right|,
$
where positive (negative) values correspond to dominant forward (backward) transmission. Strong nonreciprocal transport emerges around $\Phi\sim90^\circ$ and $\Phi\sim270^\circ$, where the phase-dependent complex coupling produces asymmetric interference between coherent and dissipative interactions. The EPs appear near the transition boundaries separating opposite isolation sectors, revealing the interplay between non-Hermitian mode coalescence and directional transport.
}
    \label{fig:Isolation_ratio}
\end{figure}

 \section{Population Dynamics}
\subsection{ Time Evolution of Mode Populations}
To investigate the dynamical evolution of the coupled photon–magnon system from an initial state toward its long-time behavior, we analyze the time-dependent populations of the photon and magnon modes under both coherently driven and undriven conditions. The temporal dynamics provide insight into the stability, transient amplification, mode hybridization, and non-Hermitian evolution of the system near the exceptional-point regime. 
Under the non-driven conditions. 
The population equation is given in this form: 
\begin{equation}
\frac{d}{dt}
\begin{pmatrix}
\langle p^\dagger p \rangle \\
\langle m^\dagger m \rangle \\
\langle p^\dagger m \rangle \\
\langle m^\dagger p \rangle
\end{pmatrix}
=
\mathrm{C}
\begin{pmatrix}
\langle p^\dagger p \rangle \\
\langle m^\dagger m \rangle \\
\langle p^\dagger m \rangle \\
\langle m^\dagger p \rangle
\end{pmatrix}.
\label{eq:matrix_dynamics}
\end{equation}
Where $\mathrm{C}$ is given by:
\begin{equation}
C=
\begin{pmatrix}
-2(\beta+\kappa) & 0 & -iG & iG \\
0 & -2(\alpha+\gamma) & iG & -iG \\
-iG & iG & -i\Delta - \kappa_T & 0 \\
iG & -iG & 0 & i\Delta - \kappa_T
\end{pmatrix}.
\label{eq:coefficient_matrix}
\end{equation}

With the detuning defined as $\Delta = \delta_m - \delta_c$ the total damping rate 
$\kappa_T =  (\beta+\kappa+ \alpha+\gamma)$, the coefficient matrix $C$ possesses four eigenvalues,
\begin{align}
\lambda_{1,2} &= -\kappa \pm \sqrt{\frac{\delta - \delta'}{2}}, \\
\lambda_{3,4} &= -\kappa \pm \sqrt{\frac{\delta + \delta'}{2}},
\label{eq:eigenvalues}
\end{align}
where
\begin{align}
\delta &= d^2 - 4G^2 - \Delta^2, \\
\delta' &= \sqrt{\delta^2 + 4\Delta^2 d^2}, \\
d &= (\beta+\kappa) - (\alpha+\gamma).
\label{eq:parameters}
\end{align}

These four eigenvalues represent four distinct decay channels of the hybrid cavity-magnon system. 
The coefficient matrix $C$ can be diagonalized through a similarity transformation
\begin{equation}
U C U^{-1} = \mathrm{diag}(\lambda_1,\lambda_2,\lambda_3,\lambda_4),
\label{eq:diagonalization}
\end{equation}
where $U$ is the eigenvector matrix.

Consequently, the dynamical equation in Eq.~(\ref{eq:matrix_dynamics}) can be transformed into independent eigenmodes,
\begin{equation}
\langle V_i(t) \rangle = \langle V_i(0) \rangle e^{\lambda_i t},
\qquad i=1,2,3,4,
\label{eq:eigenmode}
\end{equation}
which $\langle V_i(t)\rangle$ are the normal modes obtained from the average excitations
$\langle c^\dagger c\rangle$, $\langle m^\dagger m\rangle$,
$\langle c^\dagger m\rangle$, and $\langle m^\dagger c\rangle$
via the transformation matrix $U$. These hybridized modes correspond to the
cavity-magnon polaritons of the system.
\begin{equation}
   \langle V_i(t)\rangle =  \sum_{j=1}^{4} a_j e^{\lambda_jt} V_{ij} 
\end{equation}
Where $a_j = P^{-1}V(0)$ and $P^{-1}$ is eigenvector matrix and $V(0)$ defines the initial populations.

\subsection{Population dynamics under driven conditions}
The dynamical equations for the magnon and photon populations  derived in Appendix \ref{app:Population dynamics} from the Quantum Langevin equations written in matrix form as
\begin{equation}
\frac{d}{dt}
\begin{pmatrix}
\langle p^\dagger p \rangle \\
\langle m^\dagger m \rangle \\
\langle p^\dagger m \rangle \\
\langle m^\dagger p \rangle \\
\langle p^\dagger \rangle \\
\langle p \rangle \\
\langle m^\dagger \rangle \\
\langle m \rangle
\end{pmatrix}
=
D
\begin{pmatrix}
\langle p^\dagger p \rangle \\
\langle m^\dagger m \rangle \\
\langle p^\dagger m \rangle \\
\langle m^\dagger p \rangle \\
\langle p^\dagger \rangle \\
\langle p \rangle \\
\langle m^\dagger \rangle \\
\langle m \rangle
\end{pmatrix}
+
\begin{pmatrix}
0 \\
0 \\
0 \\
0 \\
\Omega_p \\
\Omega_p \\
\Omega_m \\
\Omega_m
\end{pmatrix}.
\label{General_population_matrix_equation}
\end{equation}

\begin{widetext}
\begin{equation}
D = 
\begin{pmatrix}
-2(\beta+\kappa) & 0 & -iG & iG & \Omega_p & \Omega_p & 0 & 0 \\
0 & -2(\alpha+\gamma) & iG & -iG & 0 & 0 & \Omega_m & \Omega_m \\
-iG & iG & -i\Delta-\kappa_T & 0 & \Omega_m & 0 & 0 & \Omega_p \\
iG & -iG & 0 & i\Delta-\kappa_T & 0 & \Omega_m & \Omega_p & 0 \\
0 & 0 & 0 & 0 & i\delta_p-(\beta+\kappa) & 0 & iG & 0 \\
0 & 0 & 0 & 0 & 0 & -i\delta_p - (\beta+\kappa) & 0 & -iG \\
0 & 0 & 0 & 0 & iG & 0 & i\delta_m-(\alpha+\gamma) & 0 \\
0 & 0 & 0 & 0 & 0 & -iG & 0 & -i\delta_m-(\alpha+\gamma)
\end{pmatrix}
\label{eq:D_matrix}
\end{equation}
\end{widetext}

Here, $D$ is an eight-dimensional coefficient matrix whose eigenvalues determine the eigenmode decay channels of the system. 

In addition to the four decay channels given in Eq.~(\ref{eq:eigenvalues}), the driven system contains four additional channels.
\begin{align}
\lambda_5 &= \frac{1}{2}
\left(
-\kappa + i\Delta' - \sqrt{\delta + 2i\Delta d}
\right), \\
\lambda_6 &= \frac{1}{2}
\left(
-\kappa + i\Delta' + \sqrt{\delta + 2i\Delta d}
\right),
\label{eq:driven_channels_1}
\end{align}

\begin{align}
\lambda_7 &= \frac{1}{2}
\left(
-\kappa - i\Delta' - \sqrt{\delta - 2i\Delta d}
\right), \\
\lambda_8 &= \frac{1}{2}
\left(
-\kappa - i\Delta' + \sqrt{\delta - 2i\Delta d}
\right),
\label{eq:driven_channels_2}
\end{align}
where
\begin{equation}
\Delta' = \omega_m + \omega_c - 2\omega.
\label{eq:delta_prime}
\end{equation}
The dynamical equations for the coupled photon--magnon system can be written in compact matrix form as

\begin{equation}
\frac{d}{dt}\mathbf{X}(t)=D\mathbf{X}(t)+\mathbf{F},
\end{equation}

where

\begin{equation}
\mathbf{X}(t)=
\begin{pmatrix}
\langle p^\dagger p\rangle\\
\langle m^\dagger m\rangle\\
\langle p^\dagger m\rangle\\
\langle m^\dagger p\rangle\\
\langle p^\dagger\rangle\\
\langle p\rangle\\
\langle m^\dagger\rangle\\
\langle m\rangle
\end{pmatrix},
\qquad
\mathbf{F}=
\begin{pmatrix}
0\\
0\\
0\\
0\\
\Omega_p\\
\Omega_p\\
\Omega_m\\
\Omega_m
\end{pmatrix}.
\label{XFvector}
\end{equation}
The formal time-dependent solution of Eq.~(27) is obtained as

\begin{equation}
\mathbf{X}(t)
=
e^{Dt}\mathbf{X}(0)
+
\int_{0}^{t}
e^{D(t-\tau)}
\mathbf{F}\,
d\tau,
\label{formal_solution}
\end{equation}

where $\mathbf{X}(0)$ denotes the initial state vector and
$e^{Dt}$ is the matrix exponential defined by

\begin{equation}
e^{Dt}
=
\sum_{n=0}^{\infty}
\frac{(Dt)^n}{n!}.
\end{equation}

Since the driving vector $\mathbf{F}$ is time independent,
the integral in Eq.~(\ref{formal_solution}) can be evaluated analytically,
yielding

\begin{equation}
\boxed{
\mathbf{X}(t)
=
e^{Dt}\mathbf{X}(0)
+
D^{-1}
\left(
e^{Dt}-I
\right)
\mathbf{F}
}
\label{general_solution}
\end{equation}

provided that the matrix $D$ is invertible.

If the dynamical matrix $D$ is diagonalizable, it can be expressed as

\begin{equation}
D = V\Lambda V^{-1},
\end{equation}

where $V$ is the eigenvector matrix and

\begin{equation}
\Lambda
=
\mathrm{diag}
\left(
\lambda_1,\lambda_2,\ldots,\lambda_8
\right)
\end{equation}

contains the eigenvalues of $D$.
Consequently, the matrix exponential becomes

\begin{equation}
e^{Dt}
=
V e^{\Lambda t} V^{-1},
\end{equation}

with

\begin{equation}
e^{\Lambda t}
=
\mathrm{diag}
\left(
e^{\lambda_1 t},
e^{\lambda_2 t},
\ldots,
e^{\lambda_8 t}
\right).
\end{equation}

Thus, each dynamical observable evolves as a superposition
of exponential modes,

\begin{equation}
X_k(t)
=
\sum_{j=1}^{8}
C_{kj}
e^{\lambda_j t}
+
X_k^{(p)},
\label{solution_of_8_matrix}
\end{equation}
Here, $\lambda_j$ $(j=1,2,\dots,8)$ denote the eigenvalues of the
dynamical matrix $D$, while the coefficients $C_{kj}$ are complex
constants determined by the initial conditions of the coupled
photon--magnon system through the projection of the initial state
vector onto the eigenmodes of $D$. The quantity $X_k^{(p)}$
represents the particular solution associated with the external
coherent driving fields $\Omega_p$ and $\Omega_m$, describing the
driven response of the system, independent of the transient
dynamics. Consequently, the complete time evolution consists of
a superposition of transient eigenmodes weighted by $C_{kj}$ and
the externally driven contribution $X_k^{(p)}$.

\section{SPECTRAL TOPOLOGY AS A PREDICTOR OF DYNAMICAL STABILITY}

The spectral properties discussed above provide direct insight into the dynamical behavior of the system. Because both the eigenspectrum and temporal evolution originate from the same effective non-Hermitian Hamiltonian, spectral topology and dynamical stability are fundamentally linked.

In the level-repulsion regime, coherent energy exchange dominates and all dynamical eigenvalues remain negative, ensuring stable evolution toward a steady state. As the system approaches the exceptional point, eigenvalue coalescence reduces the effective damping of the dominant hybrid mode. Beyond the exceptional point, the real part of one or more dynamical eigenvalues may become positive, leading to amplification and instability. Therefore, the transition from level repulsion to level attraction not only signifies a spectral phase transition but also acts as a precursor to dynamical instability.

\section{Stability Analysis of Population dynamics}
The stability of the coupled photon–magnon system is determined by the eigenvalue spectrum of the dynamical matrix $D$ given in the equation \eqref{eq:D_matrix}. Since the system is governed by a set of linear first-order differential equations \eqref{General_population_matrix_equation} is given by equation \eqref{solution_of_8_matrix}.

The temporal evolution of the system is controlled by the eigenmodes of the non-Hermitian matrix $D$. The corresponding eigenvalues $\lambda_i$ define the effective decay or amplification channels of the hybrid photon–magnon modes. Therefore, the sign of the real part of the eigenvalues determines the dynamical stability of the system. The phase dependence of the eigenvalue spectrum reveals three distinct dynamical regions. In Region I, all eigenvalues possess negative real parts, and the system exhibits stable dissipative evolution. In Region II, one eigenvalue approaches zero, producing critical slowing down and long-lived transient oscillations. In Region III, at least one eigenvalue acquires a positive real part, causing exponential growth of the corresponding hybrid mode. The phase parameter, therefore, acts as a dynamical control knob that continuously tunes the system between stable, critical, and unstable regimes.

The system is dynamically stable when all eigenvalues satisfy.
\[
\mathrm{Re}(\lambda_i)<0,
\]
for all $i$. In this regime, all dynamical modes decay with time, and the system evolves toward a finite steady state. The steady-state populations of the photon and magnon modes are therefore well defined.

When at least one eigenvalue satisfies
\[
\mathrm{Re}(\lambda_i)>0,
\]
The corresponding mode experiences exponential amplification. In this unstable regime, the photon and magnon populations grow exponentially with time, and the system no longer reaches a physical steady state. Such behavior originates from the competition between coherent and dissipative couplings induced by the complex interaction parameter.
\[
G=g-i\Gamma e^{i(\Theta+\phi)}.
\]
The interplay between the coherent coupling $g$, dissipative coupling $\Gamma$, and phase parameter $\phi$ gives rise to stable and unstable dynamical regions. In particular, near the exceptional point, eigenvalue coalescence and eigenvector merging lead to a strong enhancement of the system response and highly sensitive population dynamics.

The real eigenvalues of the matrix D are given as a function of $\Phi$, which can tune the stability of the system.
\begin{figure}
    \centering
    \includegraphics[width=1\linewidth]{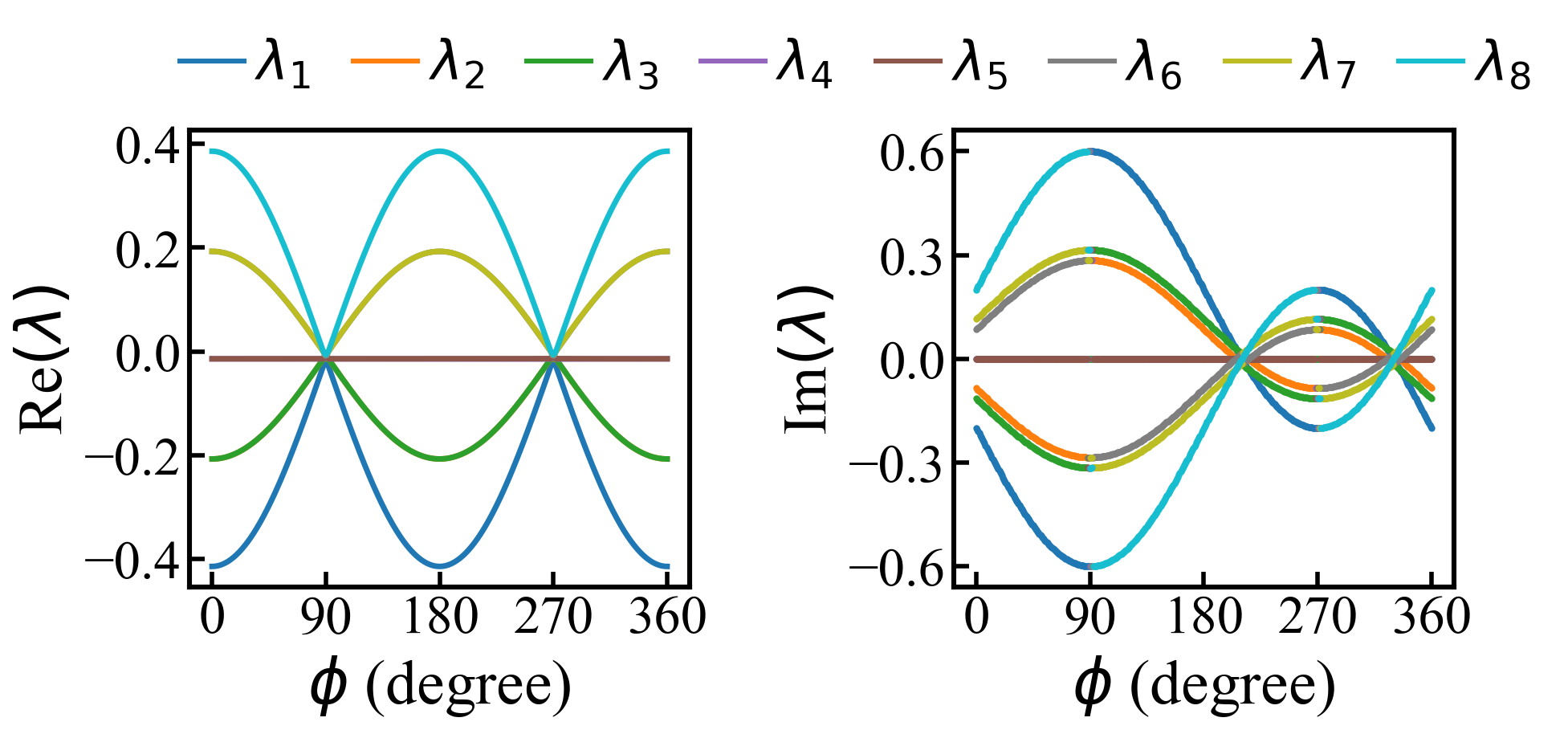}
  \caption{
Real and imaginary parts of the eigenvalue spectrum of the effective dynamical matrix as a function of the phase $\phi$. The upper panel shows the variation of $\mathrm{Re}(\lambda)$, which determines the dynamical stability of the system. Negative values of $\mathrm{Re}(\lambda)$ correspond to stable dissipative evolution, whereas positive values indicate unstable amplification of the modes. The lower panel presents the corresponding imaginary parts $\mathrm{Im}(\lambda)$, which characterize the oscillation frequencies of the hybrid photon--magnon modes and their phase-dependent evolution. The parameters used are $\kappa = 0.01$, $\beta = 0.001$, $\alpha = 0.0001$, $\gamma = 0.003$, $g = 0.06$, $\Gamma = 0.05$, $\Delta = 0$, and $\Theta = 0$.
}
    \label{fig:Eigenvalue_spectrum_of_matrix_D}
\end{figure}
 Now we begin to show the population dynamics of both Photon(cavity) and magnon modes, according to the real eigenvalues of the dynamics matrix D. 
\begin{figure}
    \centering
    \includegraphics[width=0.8\linewidth]{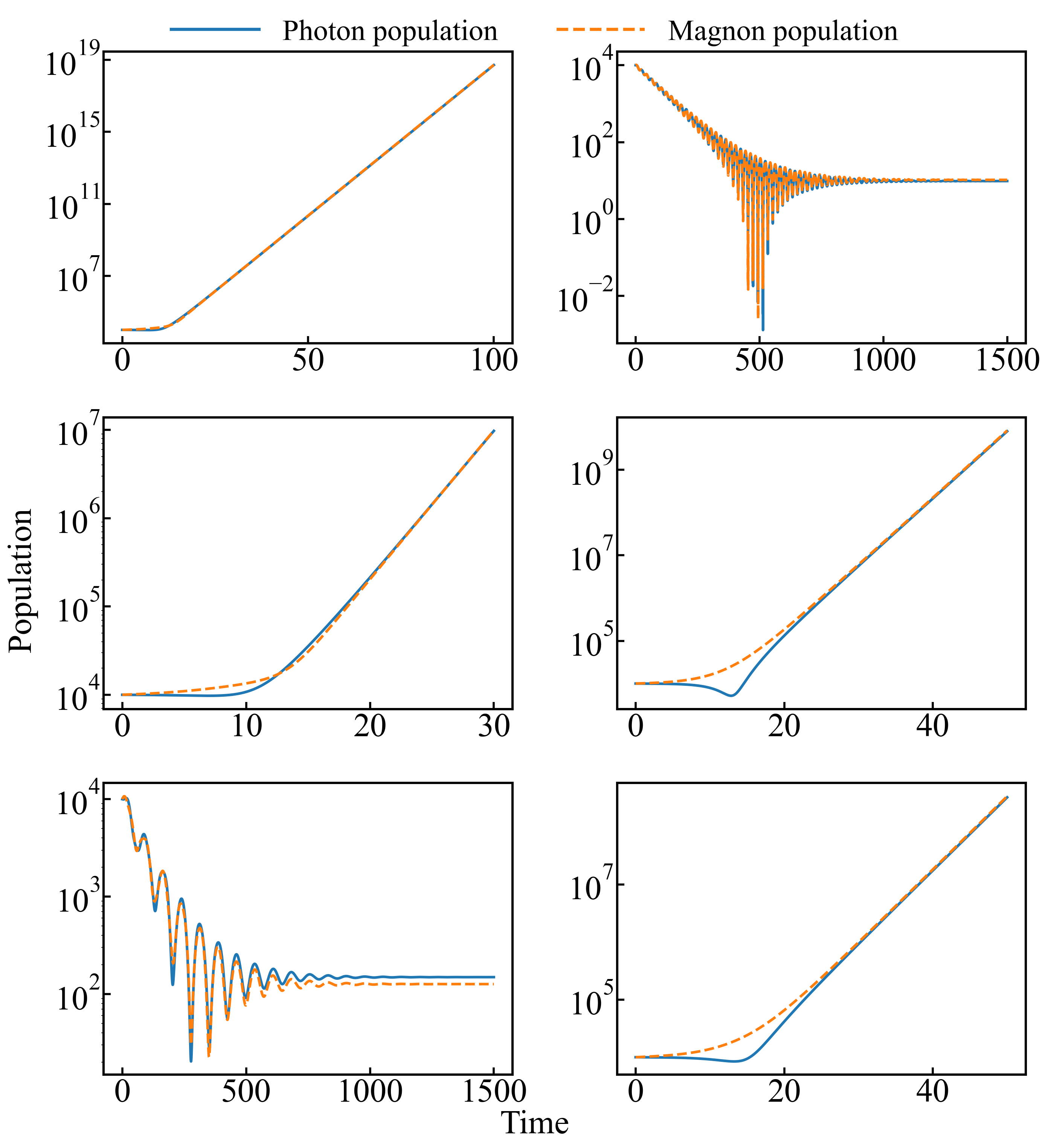}
    \caption{
Time evolution of the photon and magnon populations for
different phase values: (a) $\phi=0^\circ$,
(b) $\phi=90^\circ$, (c) $\phi=180^\circ$,
(d) $\phi=200^\circ$, (e) $\phi=270^\circ$,
and (f) $\phi=320^\circ$.
The initial photon and magnon populations are chosen as
$\langle p^\dagger p\rangle_0=
\langle m^\dagger m\rangle_0=10^4$.
The system parameters are
$\kappa=0.01$, $\beta=0.001$,
$\alpha=0.0001$, $\gamma=0.003$,
$g=0.1$, $\Gamma=0.2$,
$\delta_p=0.01$, $\delta_m=0.02$,
and $\Omega_p=\Omega_m=0.1$.
The total damping rate is defined as
$\kappa_T=\kappa+\beta+\alpha+\gamma$,
while the effective detuning is
$\Delta=\delta_m-\delta_p$.
Depending on the phase $\phi$, the system exhibits both
stable and unstable dynamical regimes.
The unstable regions are characterized by exponential
population growth associated with eigenmodes satisfying
$\mathrm{Re}(\lambda_j)>0$, whereas the stable regimes
correspond to $\mathrm{Re}(\lambda_j)<0$, leading to
decaying transient dynamics.
Furthermore, the oscillatory behavior observed in the
population evolution originates from the imaginary parts
of the eigenvalues, $\mathrm{Im}(\lambda_j)$, which
govern the coherent photon--magnon energy exchange in the
non-Hermitian hybrid system.
}
    \label{fig:population_dynamics_at_different_phi}
\end{figure}
\begin{figure*}
    \centering
    \includegraphics[width=0.72\textwidth]{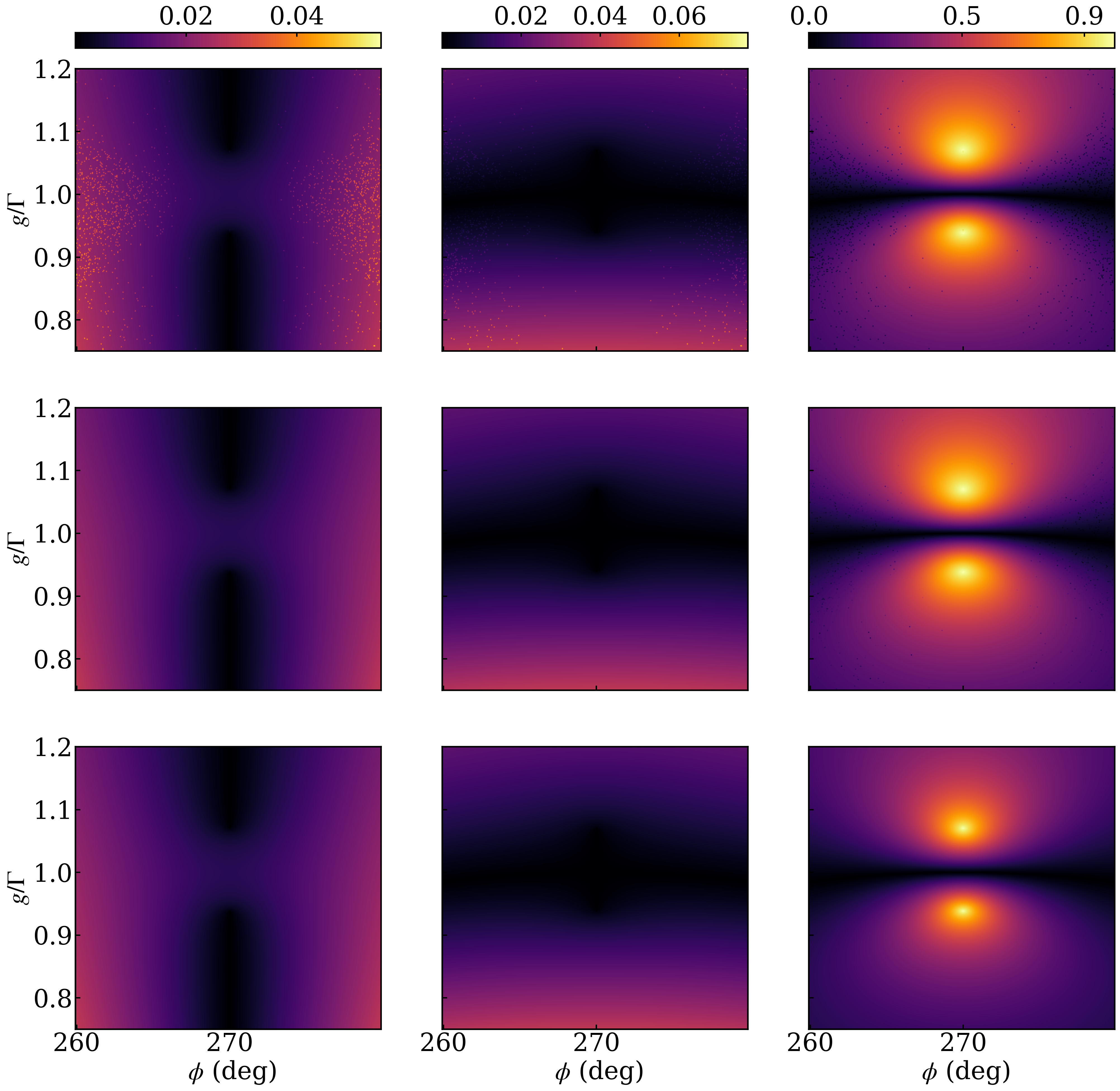}
    \caption{Phase-dependent spectral topology of the non-Hermitian photon-magnon system. Real-part splitting (a, d, g), imaginary-part splitting (b, e, h), and eigenvector overlap (c, f, i) for the eigenvalue pairs $(\lambda_1,\lambda_2)$, $(\lambda_2,\lambda_3)$, and $(\lambda_5,\lambda_6)$ versus phase $\phi$ and coupling ratio $g/\Gamma$. Regions of large overlap signify eigenstate coalescence near exceptional points. Parameters: $\kappa=0.01$, $\beta=0.001$, $\alpha=0.0001$, $\gamma=0.003$, $g=0.06$, $\Theta=0$, and $\delta_p=\delta_m=0$.}
    \label{selected_Ep_murging}
\end{figure*}
The real parts $\mathrm{Re}(\lambda_j)$ determine the
stability properties of the system, while the imaginary parts
$\mathrm{Im}(\lambda_j)$ characterize the oscillatory dynamics
associated with coherent photon--magnon energy exchange.

\section{Population Dynamics around an exceptional point}

The population dynamics provide a direct manifestation of the stability properties predicted by the eigenvalue analysis. Stable regions are characterized by bounded populations that eventually saturate. Near the exceptional point, the reduced damping leads to long-lived oscillatory behavior and enhanced energy exchange between the photon and magnon subsystems. In contrast, unstable regions exhibit exponential population growth due to the presence of dynamically amplifying eigenmodes. In the previous section, we analyzed the stability and
oscillatory behavior of the coupled photon--magnon system
through the real and imaginary parts of the eigenvalues of the dynamical matrix D. 
The observed oscillatory population transfer originates from coherent hybridization between photon and magnon modes. The oscillation frequency is governed by the imaginary parts of the dynamical eigenvalues, whereas the envelope of the oscillations is controlled by their real parts. Consequently, the phase parameter simultaneously determines the rate of energy exchange and the overall stability of the system.
\begin{figure*}
    \centering \includegraphics[width=0.82\textwidth]{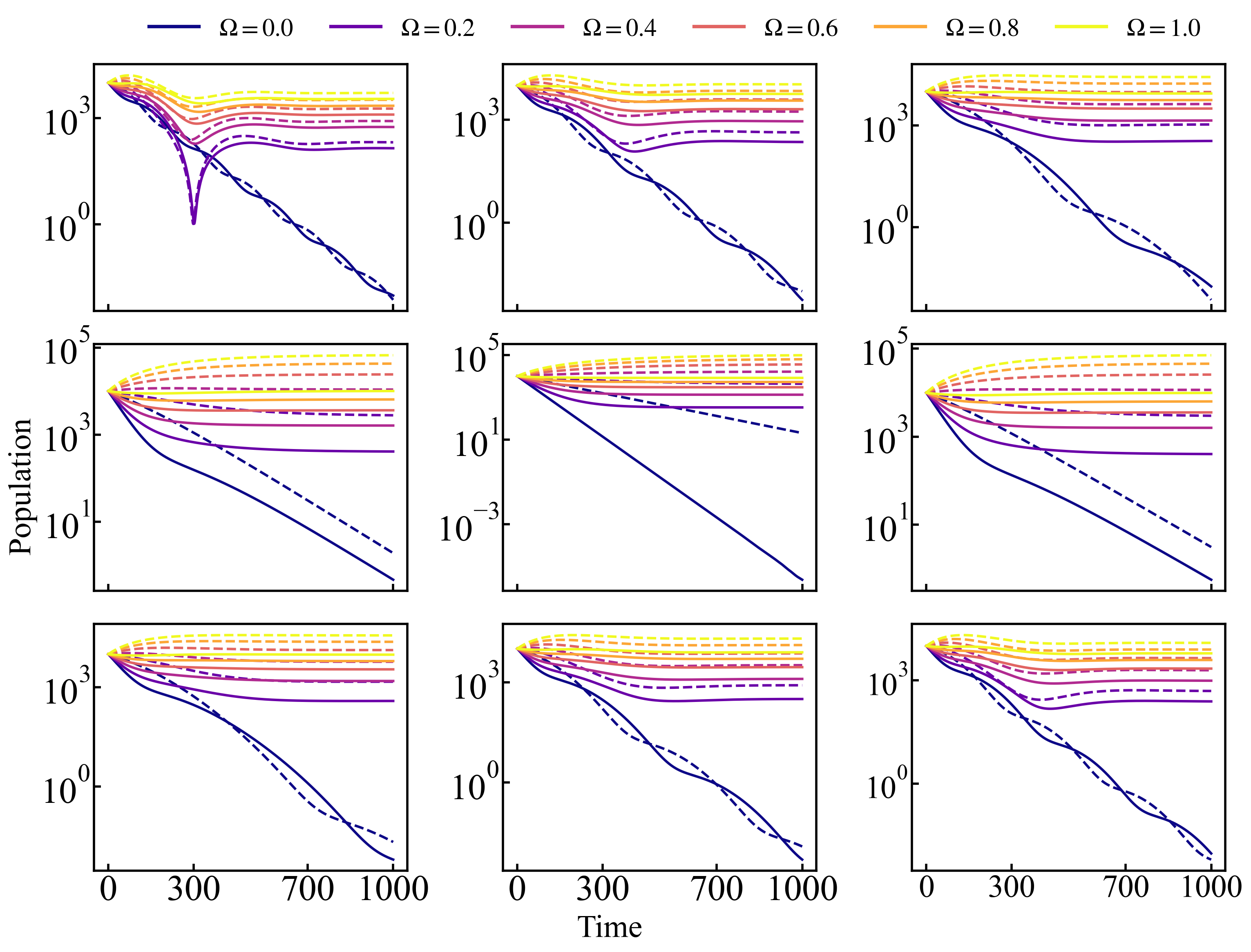}
    \caption{Photon (solid) and magnon (dotted) population dynamics for driving strengths $\Omega=0.0$–$1.0$. Panels (a)–(i) show coupling ratios $g/\Gamma=0.70$–$1.10$. Parameters: $g=0.06$, $\Phi=270^{\circ}$, $\Delta=0$, and $N_p(0)=N_m(0)=10^{4}$. Near the exceptional point ($g/\Gamma \simeq 1$), the relaxation becomes significantly slower, while stronger driving increases the steady-state populations.}
    \label{fig:population_dynamics_EP}
\end{figure*}
We now further investigate the non-Hermitian spectral properties of the system and identify the emergence of exceptional points (EPs). Exceptional points represent singular degeneracies unique to non-Hermitian systems where both eigenvalues and eigenvectors coalesce. In cavity magnonics, EPs separate qualitatively distinct interaction regimes and strongly influence transport, stability, and energy-transfer dynamics. Because a tunable phase controls the effective coupling in the present system, the location of the EPs can be manipulated externally without modifying the intrinsic properties of the cavity or magnetic material.

To identify the exceptional points, we calculate the difference between the eigenvalues as a function of the system parameters. The eigenvalue coalescence condition is given by

\begin{equation}
\Delta\lambda_{ij}
=
|\lambda_i-\lambda_j|,
\end{equation}

where $\lambda_i$ and $\lambda_j$ are the eigenvalues of the
dynamical matrix $D$. The condition
$\Delta\lambda_{ij}\rightarrow0$ indicates the merging of two
eigenmodes.

However, eigenvalue degeneracy alone does not establish the presence of an exceptional point (EP). A true EP occurs when both eigenvalues and eigenvectors simultaneously coalesce, rendering the dynamical matrix $D$ defective and non-diagonalizable. For two eigenmodes $\lambda_i$ and $\lambda_j$, the EP condition is

\begin{equation}
\lambda_i=\lambda_j,
\qquad
\mathbf{v}_i=\mathbf{v}_j,
\end{equation}

which can be numerically identified through the vanishing eigenvalue splitting,

\begin{equation}
|\lambda_i-\lambda_j| \rightarrow 0,
\end{equation}

together with the eigenvector coalescence criterion

\begin{equation}
|\langle V_i|V_j\rangle| \rightarrow 1.
\end{equation}

At this point, the matrix acquires a Jordan-block structure and cannot be fully diagonalized. Since the $8\times8$ dynamical matrix admits multiple eigenvalue coalescences, the physically relevant EPs are identified from the hybrid photon--magnon modes that dominate the observable population dynamics. To characterize the non-Hermitian phase transition, we therefore analyze the eigenvalue spectrum and eigenvector overlap as functions of the phase $\phi$ and coupling ratio $g/\Gamma$. The EP is marked by the simultaneous collapse of the real and imaginary eigenvalue splittings accompanied by maximal eigenvector overlap.
As shown in Fig.\ref{selected_Ep_murging}, the EPs emerge along phase-dependent branches in the $(\phi,g/\Gamma)$ parameter space. 
These branches separate the oscillatory regime, characterized by finite imaginary-part splitting, from the overdamped regime where the oscillation frequency vanishes. Figure 9 illustrates that the EPs form phase-dependent boundaries that separate oscillatory and overdamped dynamical regimes. Near these boundaries, the system exhibits strong mode hybridization, enhanced sensitivity to perturbations, and anomalous energy-transfer characteristics. These observations confirm that the interference phase provides a powerful means for engineering the non-Hermitian topology of cavity magnonic systems. To reveal the dynamical consequences of the exceptional-point transition, we next investigate the time evolution of the photon and magnon populations across the EP boundary. The population dynamics are evaluated at fixed phase values corresponding to the EP branches identified in the spectral phase diagram.

For each branch, the coupling ratio $g/\Gamma$ is chosen below, near, and above the EP to probe the transition between overdamped and oscillatory dynamical regimes.
The system is initialized with only magnon excitation, such that $\langle p^\dagger p\rangle_0=N_p$ and $\langle m^\dagger m\rangle_0=N_m$, allowing us to directly monitor the phase-dependent energy transfer between the magnon and photon modes.

\section{Exceptional-Point-Enhanced Sensing and Sensitivity Analysis}
Having fully characterized the steady-state population dynamics via the global $8 \times 8$ population matrix D at \eqref{eq:D_matrix}, we now independently evaluate the metrological performance of our platform using the effective $2 \times 2$ non-Hermitian Hamiltonian $H_{{eff}}$ established in Eq.\eqref{Eigenvalue_matrix}. To quantify the system's susceptibility to external target fields, we introduce an ultra-weak external perturbation $\epsilon$ acting directly as a localized signal source on the magnon subsystem. In a realistic laboratory setting, this perturbation represents a microscopic shift in the external magnetic field environment ($\delta B$), which systematically modifies the magnon resonance frequency via the gyromagnetic ratio, such that the magnon detuning is mapped as $\delta_m \rightarrow \delta_m + \epsilon$. 

To isolate the pure topological impact of the non-Hermitian singularity on sensing, we assume the system is operated on resonance ($\delta_p = \delta_m = 0$). Under this configuration, the introduction of the localized target perturbation $\epsilon$ modifies your exact eigenvalue expression from Eq.\eqref{Eigen_value_equation} into its perturbed form:
\begin{equation}
\begin{aligned}
\lambda_\pm^{(\epsilon)} = & \frac{\epsilon - i(\beta + \alpha)}{2} \\
& \pm \frac{1}{2}\sqrt{\left[ -\epsilon - i(\beta - \alpha) \right]^2 + 4G^2}.
\end{aligned}
\end{equation}

Expanding the polynomial terms inside the radical discriminant yields:
\begin{equation}
\begin{aligned}
\lambda_\pm^{(\epsilon)} ={}& \tfrac{1}{2} \big[ \epsilon-i(\beta+\alpha) \big] \\
&\pm \tfrac{1}{2} \sqrt{\epsilon^2 -i2\epsilon(\beta-\alpha)
-(\beta-\alpha)^2 + 4G^2},
\end{aligned}
\label{Perturbed_eigenvalue}
\end{equation}

\subsection{ Localization of the Exceptional Surface}
Before evaluating the sensor's responsivity, we must rigorously define the coordinates of the underlying Exceptional Points (EPs) where the complex Riemann sheets intersect \cite{Spectral_splitting_of_EPs,EPs_magnetic_senstivity_CMP}. An EP occurs precisely when the discriminant under the radical in Eq.~\eqref{Perturbed_eigenvalue} vanishes in the absence of any external perturbation ($\epsilon = 0$). Setting the unperturbed radical to zero yields the strict non-Hermitian degeneracy condition:

\begin{equation}
-(\beta - \alpha)^2 + 4G^2 = 0 \implies \frac{\beta - \alpha}{2} = \pm i G.
\label{eq:ep_condition}
\end{equation}
By expanding $G$ via Euler's formula into its real and imaginary channels, $G = [g + \Gamma\sin(\Theta+\Phi)] - i[\Gamma\cos(\Theta+\Phi)]$, Eq.~\eqref{eq:ep_condition} decouples into two distinct geometric constraints that map out an Exceptional Surface in the parameter space:
\begin{subequations}
\label{eq:exceptional_surface}
\begin{align}
\Delta_{\text{lock}} = \delta_c - \delta_m &= \pm 2\Gamma\cos(\Theta+\Phi), \\
\beta - \alpha &= \mp 2\left(g + \Gamma\sin(\Theta+\Phi)\right).
\end{align}
\end{subequations}

Equations~\eqref{eq:exceptional_surface}a and \eqref{eq:exceptional_surface}b dictate that by choosing a fixed direct coupling $g$ and dissipative coupling $\Gamma$, a laboratory user can continuously navigate along this degenerate exceptional manifold simply by sweeping the external phase shifter $\Phi$ and tracking the corresponding resonance detuning.

\begin{figure}
    \includegraphics[width=0.8\linewidth]{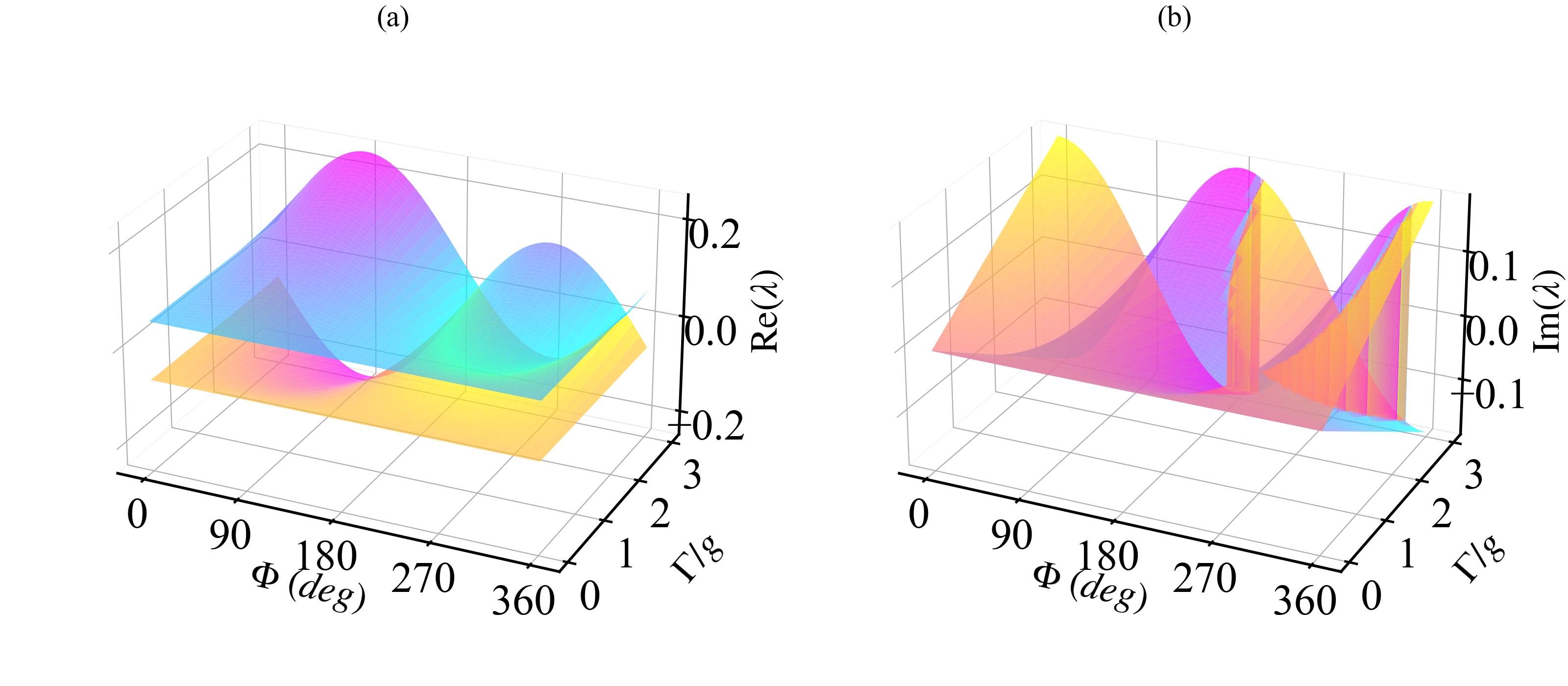}
    \caption{Real and imaginary Riemann surfaces of the eigenvalues ($\lambda_\pm$) of the effective non-Hermitian cavity--magnon Hamiltonian plotted as functions of the external phase ($\Phi$) and the normalized coupling ratio ($\Gamma/g$). (a) $\mathrm{Re}(\lambda_\pm)$ and (b) $\mathrm{Im}(\lambda_\pm)$. The phase-dependent interference between coherent and dissipative coupling channels continuously reshapes the spectral topology, producing branch-point singularities and sheet reconnections characteristic of non-Hermitian systems. The merging of the two sheets identifies the exceptional-surface manifold, where the eigenvalues and eigenvectors simultaneously coalesce. The colored surfaces correspond to the two eigenvalue branches $\lambda_+$ and $\lambda_-$. Parameters used in the calculations are $g=0.06$, $\delta_c=\delta_m=0$, $\alpha=0.0001$, $\beta=0.001$, $\Theta=0$,
}
    \label{fig:Riemann_Surface}
\end{figure}

\begin{figure}[!htbp]
    \centering
    \includegraphics[width=0.8\linewidth]{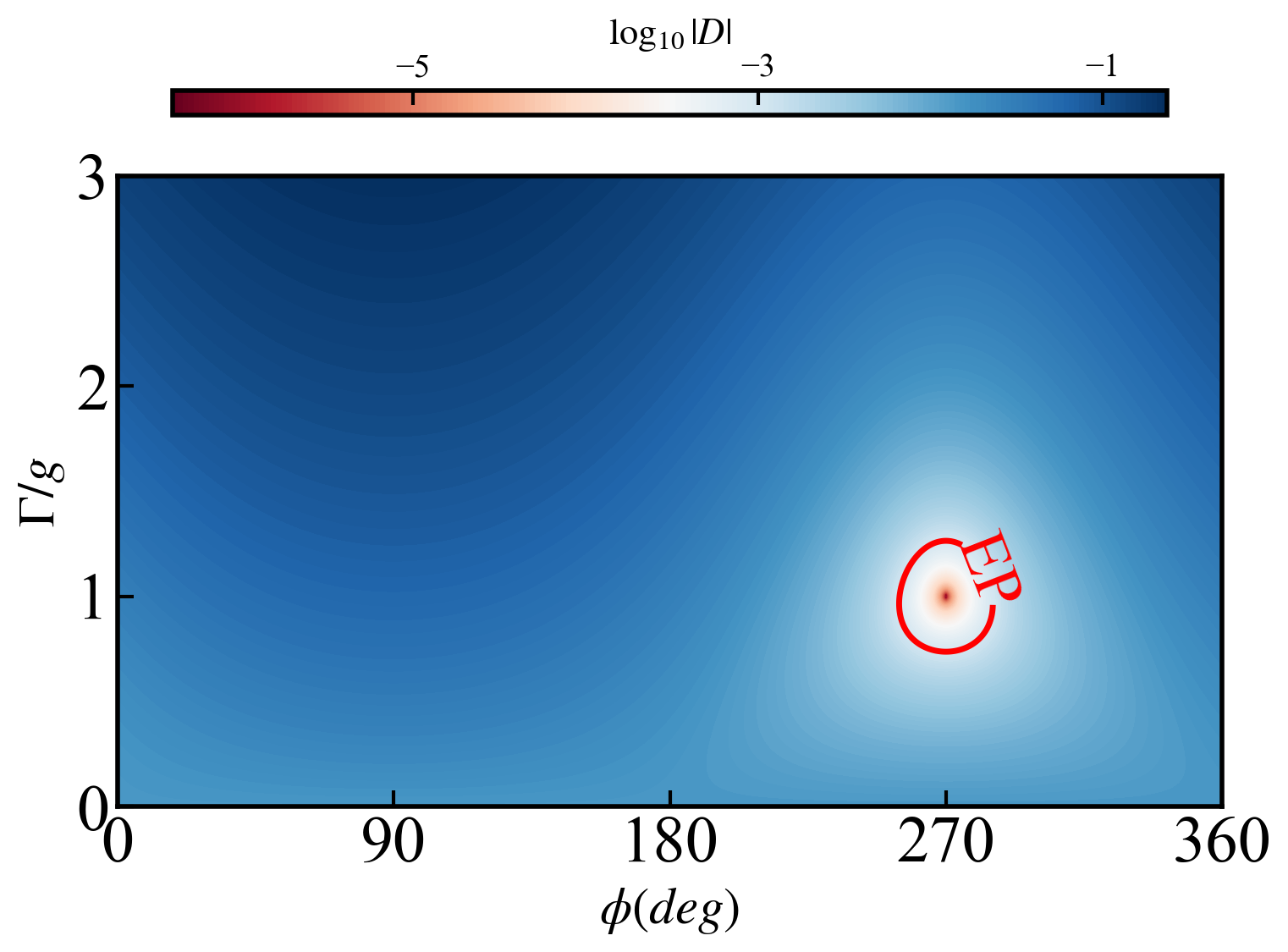}
     \caption{Colormap of $\log_{10}|D|$, where $D$ is the discriminant of the effective non-Hermitian Hamiltonian. The exceptional point is identified by the global minimum of $|D|$ in the $(\Phi,\Gamma/g)$ parameter space, indicated by the red circle. The progressive reduction of $|D|$ toward the highlighted region signifies the collapse of the eigenvalue splitting and the approach to an exceptional degeneracy. The parameters are $g=0.06$, $\delta_p=\delta_m=0$, $\alpha=0.0001$, $\beta=0.001$, and $\Theta=0$.}
    \label{fig:placeholder}
\end{figure}
\subsection{ Asymptotic Sensitivity Scaling}
When the system parameters are precisely tuned to satisfy the EP constraints in Eqs.\eqref{eq:exceptional_surface}, the unperturbed terms inside the radical cancel out exactly. To evaluate the system’s sensitivity to a weak target field ($\epsilon \ll g, \Gamma$), we execute a Taylor expansion of the discriminant in Eq.~\eqref{Perturbed_eigenvalue} around the EP coordinates. Retaining only the dominant first-order term in $\epsilon$, the eigenvalue splitting $\Delta\lambda = \lambda_+ - \lambda_-$ reduces asymptotically to:

\begin{equation}
\Delta\lambda \approx \sqrt{-2i\epsilon(\beta - \alpha)} = 2\sqrt{\pm i G\epsilon}.
\label{eq:scaling_law}
\end{equation}

\begin{figure}[!htbp]
    \centering
    \includegraphics[width=0.9\linewidth]{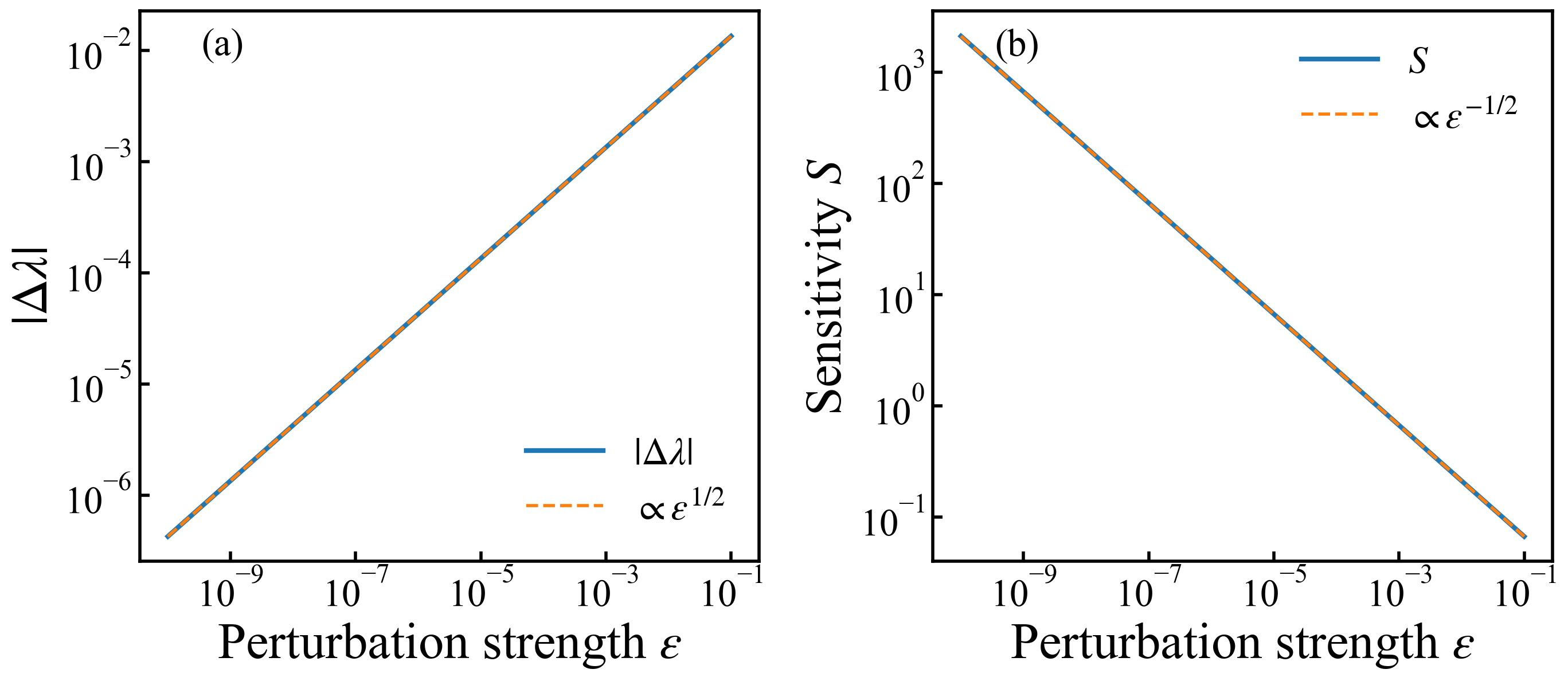}
    \caption{\textbf{Exceptional-point sensing characteristics.} (a) Eigenvalue splitting $|\Delta\lambda|$ as a function of perturbation strength $\epsilon$, following the scaling $|\Delta\lambda|\propto\epsilon^{1/2}$. (b) Responsivity $S$, exhibiting $S\propto\epsilon^{-1/2}$. The fitted slopes agree with the analytical exceptional-point scaling laws.}
    \label{fig:ep_sensing_characteristics}
\end{figure}
The mathematical responsivity of the sensor, defined as the susceptibility metric $S = |\partial(\Delta\lambda)/\partial\epsilon|$, scales as $S \propto 1/\sqrt{\epsilon}$ \cite{Lau2018}. As the target perturbation approaches the infinitesimal limit ($\epsilon \to 0$), the sensitivity slope diverges toward infinity ($S \to \infty$). This radical square-root scaling law provides an order-of-magnitude amplification over conventional, linear Hermitian sensors ($\Delta\lambda_{\text{normal}} \propto \epsilon$), demonstrating the metrological superiority of our cavity magnonic platform \cite{Enhancing_senstivity_using_EPs}.

Unpacking the complex coupling $G$ reveals how the phase sweep actively modulates the sensor's raw output response:

\begin{equation}
\Delta\lambda \propto \left( \left[g + \Gamma\sin(\Theta+\Phi)\right]^2 + \left[\Gamma\cos(\Theta+\Phi)\right]^2 \right)^{1/4} \cdot \sqrt{\epsilon}.
\end{equation}

\begin{figure}[!htbp]
    \centering
    \includegraphics[width=0.8\linewidth]{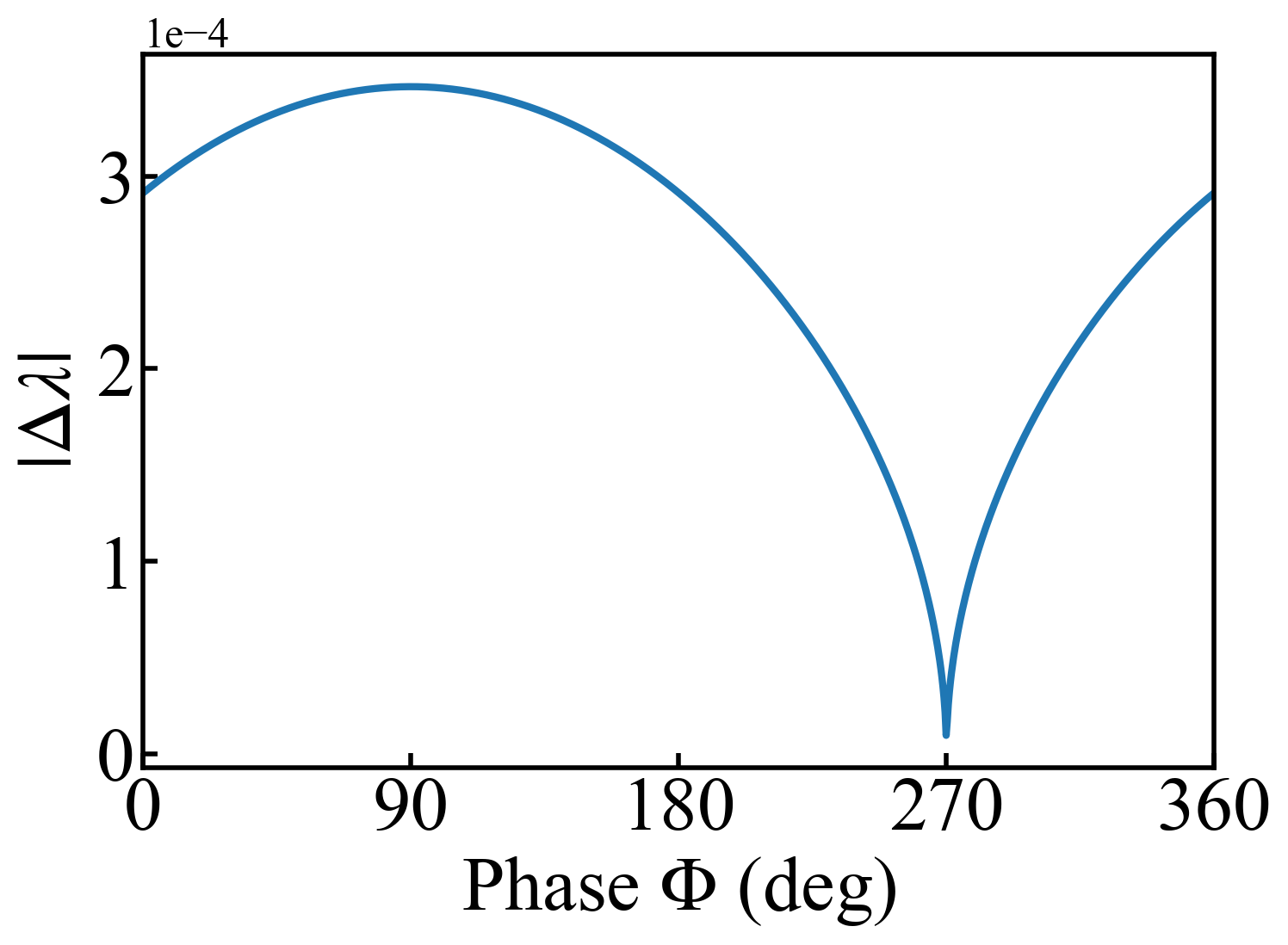}
   \caption{\textbf{Phase-controlled eigenvalue splitting near the exceptional point.} The splitting magnitude $|\Delta\lambda|$ versus propagation phase $\Phi$ for $\epsilon=10^{-6}$ and $\Gamma/g=1$. The minimum at $\Phi=270^\circ$ marks the exceptional-point condition and exhibits the characteristic square-root sensitivity scaling $|\Delta\lambda|\propto\sqrt{\epsilon}$.}

    \label{fig:phase_dependent_splitting}
\end{figure}

\subsection{Fundamental Operational Noise Floor and Power-Spectrum Scaling Laws}

While the theoretical sensing susceptibility $S$ derived via the eigenvalue splitting exhibits an asymptotic divergence at the critical threshold, a comprehensive metrological evaluation requires assessing the Signal-to-Noise Ratio (SNR) and operational resolution boundaries to account for concurrent noise amplification near non-Hermitian singularities~\cite{Enhancing_senstivity_using_EPs}. In an experimental cavity-magnonics architecture, the fundamental tracking precision is intrinsically bounded by room-temperature thermal fluctuations and intrinsic subsystem decay channels. We model the effective noise spectral density $S_N(\omega)$ entering through the cavity-magnon ports via input-output relations. 
Due to eigenvector coalescence and nonorthogonality at the exceptional point, the system exhibits Petermann-factor-induced noise enhancement. For a given laboratory measurement integration time $\tau$, the overall power-spectrum SNR is formally expressed as:

\begin{equation}
\text{SNR} = \frac{4|G(\Phi)|\tau}{S_N(\omega)\kappa_{\text{eff}}} \cdot \epsilon,
\end{equation}

where $\kappa_{\text{eff}}$ represents the effective linewidth of the coalesced polariton mode broadened by the Jordan-block structure. The complex coupling $G(\Phi) = g - i\Gamma e^{i(\Theta+\Phi)}$ encodes the phase-tunable interference between coherent and dissipative channels.

\begin{figure*}
\centering
\includegraphics[width=0.8\textwidth]{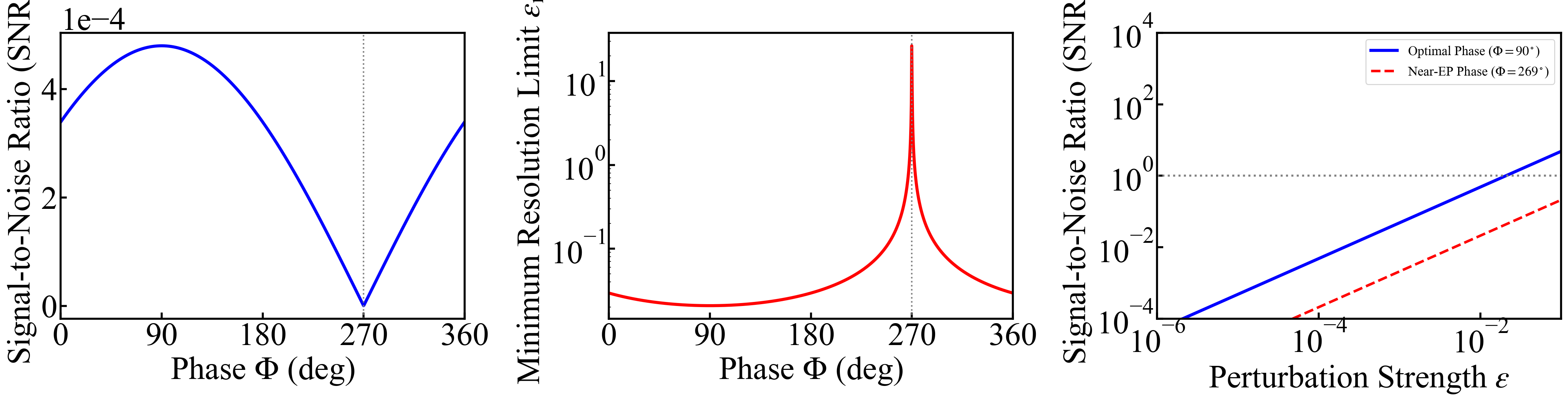}
\caption{Metrological performance and resolution limits of the phase-tunable cavity--magnonics system. (a) Signal-to-noise ratio (SNR) as a function of the propagation phase $\Phi$ for a fixed perturbation $\epsilon=10^{-5}$, showing strong degradation near the singularity. (b) Minimum detectable perturbation $\epsilon_{\mathrm{min}}$ versus $\Phi$ on a semi-log scale. At the exceptional-point crossing ($\Phi=270^\circ$), cancellation of the effective coupling leads to a divergence of the noise floor ($\epsilon_{\mathrm{min}}\rightarrow\infty$). (c) Log-log scaling of SNR with perturbation strength $\epsilon$ for the optimal phase ($\Phi=90^\circ$) and a phase near the singularity ($\Phi=269^\circ$). The horizontal dashed line indicates the detection threshold ($\mathrm{SNR}=1$). Both cases exhibit linear scaling ($\mathrm{SNR}\propto\epsilon$), demonstrating that practical sensing resolution remains limited by linear response despite non-Hermitian enhancement.}
\label{fig:unified_snr_analysis}
\end{figure*}

Evaluating this relation reveals two crucial physical characteristics of the non-Hermitian metrology setup:

First, we analyze the phase-dependent features under a fixed target perturbation ($\epsilon = 10^{-5}$). As illustrated in Fig.~\ref{fig:unified_snr_analysis}(a), the destructive interference between the coherent and dissipative coupling paths leads to a cyclic modulation of the extractable SNR, dropping identically to zero at the critical crossing ($\Phi = 270^\circ$). By establishing the fundamental tracking resolution boundary at the unity performance threshold ($\text{SNR} = 1$), the minimum detectable external perturbation $\epsilon_{\text{min}}$ under the phase constraint is strictly governed by:
\begin{equation}
\epsilon_{\text{min}}(\Phi) = \frac{S_N(\omega)\kappa_{\text{eff}}}{4|G(\Phi)|\tau} = \frac{S_N(\omega)\kappa_{\text{eff}}}{4g\tau \sqrt{2(1+\sin\Phi)}}.
\end{equation}
As mapped on the semi-logarithmic vertical scale in Fig.~\ref{fig:unified_snr_analysis}(b), this resolution boundary exposes the ultimate trade-off of the platform. At the exact exceptional point crossing ($\Phi = 270^\circ$), the absolute signal power is completely absorbed ($|G| \rightarrow 0$), forcing an asymptotic divergence in the operational noise floor ($\epsilon_{\text{min}} \rightarrow \infty$).

Second, we isolate the scaling behavior of the readout signal relative to the perturbation strength $\epsilon$. As displayed on the log-log scaling plot in Fig.~\ref{fig:unified_snr_analysis}(c), the extracted power-spectrum SNR scales strictly linearly with the target perturbation strength ($\text{SNR} \propto \epsilon^1$), which remains invariant when shifting from the optimal constructive configuration ($\Phi = 90^\circ$) to a phase window close to the singularity valley ($\Phi = 269^\circ$). This linear dependence represents an unyielding metrological boundary near non-Hermitian branch points, where Petermann noise scaling cancels out the sub-linear eigenvalue enhancement ($\mathcal{O}(\sqrt{\epsilon})$). Consequently, optimizing the device requires a strategic compromise: engineering larger base coherent coupling parameters $g$ to lower the overall boundary floor, while operating in close proximity to, rather than exactly at, the singularity vertex to exploit the high susceptibility without losing the signal to the noise floor.

\section{Conclusion}
In summary, we have developed a unified theoretical framework for controlling spectral topology, nonreciprocal transport, exceptional-point physics, and dynamical stability in non-Hermitian cavity magnonic systems through reservoir-engineered interference. By introducing a phase-dependent complex coupling arising from the interplay between coherent and dissipative interaction channels, we demonstrate that a single externally tunable phase continuously transforms the system between coherent-dominated and dissipative-dominated regimes.

The interference phase governs the transition between level repulsion and level attraction, determines the formation and location of exceptional points, and produces strongly asymmetric forward and backward transport through direction-dependent hybridization. Furthermore, the same phase controls the dynamical stability of the coupled photon–magnon system, enabling transitions between stable relaxation, critical slowing down, oscillatory energy exchange, and exponential amplification.

Our results reveal that spectral topology and dynamical behavior are two manifestations of the same underlying phase-controlled non-Hermitian interaction. The demonstrated ability to manipulate these phenomena through a single external control parameter establishes reservoir-engineered interference as a versatile tool for designing reconfigurable cavity magnonic devices. The proposed framework may enable future developments in nonreciprocal microwave components, exceptional-point-enhanced sensing, dynamically tunable hybrid quantum systems, and programmable non-Hermitian magnonic networks.

\section*{Acknowledgments}

This work was supported by the Science and Engineering Research Board (SERB), India (Grant No. SRG / 2023 / 001355), and the Anusandhan National Research Foundation (ANRF), India
(Sanction Nos. ANRF / IRG / 2025 / 001896 / PS and ANRF / ARG / 2025 / 006596 / PS). Additional support was received from the Council of Science and Technology, Uttar Pradesh (CSTUP) under Project IDs 2470 (Sanction No. CST/D - 1520) and 4482 (Sanction No. CST/D - 7/8). S. Verma acknowledges the Ministry of Education, Government of India, for the Prime Minister’s Research
Fellowship (PMRF ID-1102628). 

\section*{Data Availability}

The data that support the findings of this study are available within the article.

\appendix
\section{Derivation of the Rotating wave approximation(RWA) Hamiltonian}
\label{app:RWA_hamiltonian}
\renewcommand{\theequation}{A\arabic{equation}}
\setcounter{equation}{0}
We can simplify the time-dependent dynamics and isolate the physically relevant resonant process in this system by using the rotating wave approximation (RWA). We begin with the  Hamiltonian in the Schrödinger picture, which includes the free photon and magnon modes. This full Hamiltonian captures all the relevant dynamical operations in the hybrid quantum system (cavity magnonics) composed of one ferromagnetic magnon mode inside a microwave cavity, driven by an external field. The Hamiltonian takes the following form:
\begin{align}
\hat{H}(t)
&= \hbar\omega_p \hat{p}^\dagger(t)\hat{p}(t)
 + \hbar\omega_m \hat{m}^\dagger(t)\hat{m}(t)
\nonumber\\
&\quad
 + \hbar g
 \bigl(\hat{p}^\dagger(t)+\hat{p}(t)\bigr)
 \bigl(\hat{m}^\dagger(t)+\hat{m}(t)\bigr).
\end{align}
To derive the rotating-wave approximation (RWA), we move to the interaction picture defined with respect to the free Hamiltonian, which describes the uncoupled photon and magnon modes. The free and interaction Hamiltonians are given by:

\begin{align}
\hat{H}_{01}(t) &= \hbar \omega_p \hat{p}^\dagger \hat{p}
+ \hbar \omega_m \hat{m}^\dagger \hat{m}, \nonumber \\
\hat{H}_I(t) &= \hbar g (\hat{p}^\dagger + \hat{p})(\hat{m}^\dagger + \hat{m}).
\end{align}

To eliminate the explicit time dependence in the system Hamiltonian and facilitate the RWA, we transform this Hamiltonian into the rotating frame defined by a suitable unitary transformation. Let's introduce a unitary operator U(t) that maps a state $\ket{\psi}$ in the Schrödinger picture to a state $\ket{\psi_r}$ in the rotating frame with angular frequency $\omega$. In the rotating frame, we consider ($\hbar = 1$ throughout)\cite{Roccati2022NonHermitian}. We define $U(t) = e^{it\hat{H}_0(t)}$. Using this transformation, the $\ket{\psi_r}$ given by 
\begin{align*}
    \ket{\psi_r}= U(t)\ket{\psi}.
\end{align*} By taking the time derivative of these transformation equations of state, leads to 
\begin{align*}
    i\frac{d}{dt}\ket{\psi_r}=(-\hat{H}_{01}(t) + e^{it\hat{H}_0}\hat{H}_Ie^{-it\hat{H}_0})\ket{\psi_r}
\end{align*}.
Where, $\hat{H}_{0}(t) = \omega(\hat{p}^\dagger(t)  \hat{p}(t)  
+  \hat{m}^\dagger(t)  \hat{m}(t))$
The Hamiltonian governing the dynamics of the rotating frame is 
$\hat{H}_r(t)=(-\hat{H}_{01}(t) + e^{it\hat{H}_0}\hat{H}_I(t)e^{-it\hat{H}_0} $. This transformation allows isolating the time-independent resonant terms in $\hat{H}_r(t)$, which are preserved under the RWA. Now, proceed to compute the time evolution of the interaction picture using the transformation $e^{it\hat{H}_0}\hat{H}_I(t)e^{-it\hat{H}_0}$. To compute this, we use the Baker-Campbell-Hausdorff identity and the commutation relation of bosonic operators. 
\begin{align*}
[\hat{c}_i, \hat{c}_j^\dagger]
&= \delta_{ij}, \quad
[\hat{c}, \hat{c}] = 0, \quad
[\hat{c}^\dagger, \hat{c}^\dagger] = 0, \\
[\hat{c}^\dagger \hat{c}, \hat{d}]
&= -\hat{d}, \quad
[\hat{c}^\dagger \hat{c}, \hat{d}^\dagger] = \hat{d}^\dagger, \\
e^{i\hat{H}_0 t}(\hat{A}(t)\hat{B}(t))e^{-i\hat{H}_0 t}
&=
\Bigl(
e^{i\hat{H}_0 t}\hat{A}(t)e^{-i\hat{H}_0 t}
\Bigr)
\nonumber\\
&\quad\times
\Bigl(
e^{i\hat{H}_0 t}\hat{B}(t)e^{-i\hat{H}_0 t}
\Bigr), \\
e^{i\hat{H}_0 t}\hat{A}(t)e^{-i\hat{H}_0 t}
&=
\hat{A}e^{-i\omega_a t}, \\
e^{i\hat{H}_0 t}\hat{A}^{\dagger}(t)e^{-i\hat{H}_0 t}
&=
\hat{A}^{\dagger}e^{i\omega_a t}.
\end{align*}
For the commutation relation, where $\hat{c}, \hat{d} \in \{\hat{p}, \hat{m}\}$.  should be different.
where $\hat{A}, \hat{B} \in \{\hat{p}, \hat{p}^\dagger, \hat{m}, \hat{m}^\dagger\}$, $\omega_a$ is the frequency associated with the operator.

The Hamiltonian modified in the rotating frame is
\begin{align}
\hat{H}_r(t) ={}& -\hat{H}_0(t)+\hat{H}_{01}(t) \nonumber \\
& +g \Big( \hat{p}^\dagger \hat{m}^\dagger e^{i(\omega_p+\omega_m)t} + \hat{p}\hat{m}e^{-i(\omega_p+\omega_m)t} \nonumber \\
& \quad + \hat{p}^\dagger\hat{m}e^{i(\omega_p-\omega_m)t} + \hat{p}\hat{m}^\dagger e^{-i(\omega_p-\omega_m)t} \Big) \nonumber \\
& +i\sqrt{\kappa}P_{\mathrm{in}} \Big( \hat{p}^\dagger e^{i(\omega+\omega_p)t} - \hat{p}e^{-i(\omega_p-\omega)t} \nonumber \\
& \quad + \hat{p}^\dagger e^{i(\omega_p-\omega)t} - \hat{p}e^{-i(\omega_p+\omega)t} \Big) \nonumber \\
& +i\sqrt{\gamma}P_{\mathrm{in}} \Big( \hat{m}^\dagger e^{i(\omega+\omega_m)t} - \hat{m}e^{-i(\omega_m-\omega)t} \nonumber \\
& \quad + \hat{m}^\dagger e^{i(\omega_m-\omega)t} - \hat{m}e^{-i(\omega_m+\omega)t} \Big).
\label{eq:Hamiltonian_split}
\end{align}

Now applying RWA to the Hamiltonian $\hat{H}_r(t)$, $\hat{H} = \lim_{T \to \infty} \frac{1}{T} \int_{0}^{T} \hat{H}_r(t)\, dt$. In applying RWA, we know that $\hat{H_0}(t) =\hat{H_0}$,  $\hat{H}_{01}(t) =\hat{H}_{01}$ in the rotating frame. 
Mathematically, for a sufficiently large evolution time $T$, any fast oscillating term of the form $e^{\pm i \Omega t}$, $\Omega = (\omega+\omega_a)$ satisfies
\begin{equation}
\lim_{T \to \infty} \frac{1}{T} \int_{0}^{T} e^{\pm i \Omega t} \, dt = 0 \quad \text{for large } \Omega.
\label{RWA_condition_1}
\end{equation}

However, when the exponent arises from a near-resonant interaction such as $\omega_p - \omega_a \approx 0$, the approximation must be applied with care. In such cases, the integral becomes

\begin{equation}
\lim_{T \to \infty} \frac{1}{T} \int_{0}^{T} e^{i(\omega_p - \omega_a)t} \, dt
= \lim_{T \to \infty} \frac{1}{T} \left( \frac{e^{i(\omega_p - \omega_a)T} - 1}{i(\omega_p - \omega_a)} \right).
\label{RWA_conditions_2}
\end{equation}

Although the expression does not strictly vanish, its prefactor remains finite as the denominator becomes small when the frequency difference approaches zero. Hence, such terms are retained. Under the rotating wave approximation (RWA), rapidly oscillating terms $e^{\pm i(\omega_p+\omega_m)}$, which average to zero over long times, are neglected, yielding the RWA Hamiltonian.
\begin{align}
\hat{H}_{\mathrm{RWA}}
&=
\delta_p \hat{p}^\dagger \hat{p}
+\delta_m \hat{m}^\dagger \hat{m}
\nonumber\\
&\quad
+g\left(
\hat{p}^\dagger \hat{m}
+\hat{m}^\dagger \hat{p}
\right)
\nonumber\\
&\quad
+i\sqrt{\kappa}\,p_{\mathrm{in}}
(\hat{p}^\dagger-\hat{p})
\nonumber\\
&\quad
+i\sqrt{\gamma}\,p_{\mathrm{in}}
(\hat{m}^\dagger-\hat{m}) .
\end{align}
where $\delta_p = \omega_p -\omega$ and $\delta_m = \omega_m - \omega$.
\section{Dissipation and damping effect}   
\label{app:dissipation and damping in H RWA}
In realistic quantum systems, coupling to environmental degrees of freedom inevitably induces dissipation and decoherence. The dynamics of the system are governed by the von Neumann-Lindblad equation \cite{breuer2002theory},
\begin{equation}
\frac{d\hat{\rho}}{dt}
= -i [\hat{H}_{\mathrm{RWA}}, \hat{\rho}]
+ \mathcal{L}[\hat{\rho}],
\label{von Neumann-Lindblad equation}
\end{equation}
which captures both coherent and dissipative magnon-photon interactions. Where $\hat{H}_{RWA}$ is the coherent  Hamiltonian of the system that describes the energy of relevant modes and the coherent coupling between them, it is derived in Appendix (\ref{app:RWA_hamiltonian}).

The standard dissipative superoperator $\mathcal{L}[\hat{o]}$ is defined as:
\begin{equation}
    \mathcal{L}[\hat{o}] = \hat{o}\hat{\rho}\hat{o}^\dagger -\frac{1}{2}\hat{o}^\dagger\hat{o}\hat{\rho}-\frac{1}{2}\hat{\rho}\hat{o}^\dagger\hat{o}
\end{equation}
The jump operator has the general form\cite{Nonreciprocal_Photon}:
\begin{equation}
\hat{o} \to \lambda_p\,\hat{p} + \lambda_m\,\hat{m}.
\end{equation}
Here, the coefficients $\lambda_p$ and $\lambda_m$ characterize the individual couplings of cavity photons and magnons to traveling waves, which leads to dissipative photon-magnon coupling.
The system dynamics are governed by
\begin{equation}
\frac{d\hat{\rho}}{dt}
= -i [\hat{H}_{\mathrm{coh}}, \hat{\rho}]
+ \tau\,\mathcal{L}[\hat{o}]\,\hat{\rho}
+ \beta\,\mathcal{L}[\hat{a}]\,\hat{\rho}
+ \alpha\,\mathcal{L}[\hat{b}]\,\hat{\rho},
\end{equation}
where it $\mathcal{L}[\hat{o}]$ captures the cooperative dissipation,
while $\mathcal{L}[\hat{a}]$ and $\mathcal{L}[\hat{b}]$ describe intrinsic
losses. The induced decay rates satisfy
$\tau \lambda_p^2 = \kappa$,
$\tau \lambda_m^2 = \gamma$,
and $\tau \lambda_p \lambda_m = \sqrt{\kappa\gamma}$.
So the equation of motion for the cavity mode $\hat{p}$ and the magnon
mode $\hat{m}$ can be given as:

\textbf{For the photon mode:}
\begin{equation}
    \frac{d}{dt}\hat{p}= -i\delta_p \hat{p}-(\beta+\kappa)\hat{p} - (ig+\sqrt{\kappa\gamma})\hat{m}
    \label{Appendix_dynamics_of_photon}
\end{equation}
\textbf{For the magnon mode:}
\begin{equation}
     \frac{d}{dt}\hat{m}= -i\delta_m \hat{p}-(\alpha+\gamma)\hat{m} - (ig+\sqrt{\kappa\gamma})\hat{p}
     \label{Appendix_dynamics_of_magnon}
\end{equation}
From the above relations, the effective dissipative coupling between the cavity and magnon modes is identified as $-i\sqrt{\kappa\gamma}$. Including the coherent interaction, the total coupling takes the form $g - i\Gamma$, where $\Gamma \equiv \sqrt{\kappa\gamma}$. 
Upon reversing the propagation direction of the microwave field, the jump operator transforms as $\hat{o} \to \lambda_p \hat{a} - \lambda_m \hat{b}$, resulting in a sign reversal of the dissipative contribution, such that the effective coupling becomes $g + i\Gamma$. This change in propagation direction effectively introduces a phase shift $\pi$ in the system.

More generally, the propagation-induced phase can be treated as a tunable parameter. By incorporating a phase shifter, one can introduce a controllable phase $\Phi $, thereby continuously tuning the coupling between the cavity and magnon modes. In this way, the direction of propagation provides a direct means to control the phase of the dissipative interaction.
For forward (backward) propagation, the effective coupling is
$g - i\,\Gamma e^{i\Phi} 
 \left( g + i\,\Gamma e^{i(\Phi)} \right).$

Within the rotating wave approximation, we obtain a general effective non-Hermitian Hamiltonian for both forward and backward propagation in the coupled system:

 \begin{equation}
\hat{H}
= \tilde{\delta}_p \hat{p}^\dagger \hat{p}
+ \tilde{\delta}_m \hat{m}^\dagger \hat{m}
+ \left( g - i\,\Gamma e^{i(\Theta+\Phi)} \right)
\left( \hat{p}^\dagger \hat{m} + \hat{p}\hat{m}^\dagger \right)
\end{equation}

For the forward and backward directions, $\Theta$ is 0 and $\pi$, respectively.
where $\tilde{\delta}_p$ and $\tilde{\delta}_m$ denote the complex eigenfrequencies of the uncoupled cavity and magnon modes, respectively, given by $\tilde{\delta}_p = \delta_c - i\beta$ and $\tilde{\delta}_m = \delta_m - i\alpha$.

\section{Transmission Coefficient: Input-Output Theory}
\label{app:Transmission Coefficient}
To explore the spectral response of the system, we analyze the dynamics in the frequency domain. Starting from the Hamiltonian in the rotating frame under the rotating wave approximation (RWA), we derive the Heisenberg–Langevin equations for the system operators.

Assuming harmonic time dependence and focusing on the steady-state response, the Heisenberg–Langevin equations reduce to algebraic relations in the frequency domain.

\textbf{For the photon mode:}
\begin{equation}
    \frac{d}{dt}\hat{p}= -i\delta_p \hat{p}-(\beta+\kappa)\hat{p} - ({g} - i\Gamma e^{i(\Theta+\Phi)})\hat{m}+\sqrt{\kappa}p_{in}
    \label{dynamics_of_photon}
\end{equation}
\textbf{For the magnon mode:}
\begin{equation}
     \frac{d}{dt}\hat{m}= -i\delta_m \hat{m}-(\alpha+\gamma)\hat{m} - ({g} - i\Gamma e^{i(\Theta+\Phi)})\hat{p}+\sqrt{\gamma}p_{in}
     \label{dynamics_of_magnon}
\end{equation}

In this regime, the system reaches a time-independent solution,($\frac{d\hat{p}}{dt}$= $\frac{d\hat{m}}{dt}$ = 0). Consequently, Eqs.~\eqref{dynamics_of_photon} and \eqref{dynamics_of_magnon} reduce to a set of coupled algebraic equations. Solving these equations enables us to express the photon mode $\hat{p}$ and the magnon mode $\hat{m}$ in terms of the model parameters.
\begin{equation}
A_p\hat{p}= -G \hat{m}+\sqrt{\kappa}p_{in}
\label{steady_state_eq_1}
\tag{4.a}
\end{equation}
\begin{equation}
A_m \hat{m}= -G \hat{p}+\sqrt{\gamma}p_{in}
\label{steady_state_eq_2}
\tag{4.b}
\end{equation}
The following, after solving Eq. \eqref{steady_state_eq_2} for its $\hat{m}$ and substituting it into Eq. \eqref{steady_state_eq_1} we get 
\begin{align*}
\hat{p} &=
\frac{-A_m \sqrt{\kappa} - G \sqrt{\gamma}}
{A_p A_m - G^2} \, p_{\mathrm{in}} \\
\\
\hat{m} &=
\frac{-G \sqrt{\kappa} - A_p \sqrt{\gamma}}
{A_p A_m - G^2} \, p_{\mathrm{in}}
\end{align*}
Where, $A_p = i\delta_p + (\beta + \kappa),
A_m = i\delta_m + (\alpha + \gamma),
G = g - i\Gamma e^{i(\Theta+\Phi)}$.

To analyze the spectral and transport properties of the coupled system, we adopt an open-system description by coupling the cavity mode to external channels. Within this framework, input-output theory provides a systematic method to compute the transmission response from the internal dynamics. \ gives the relation between the input and output fields 
\begin{align}
\hat{p}_{\mathrm{out}} = \hat{p}_{\mathrm{in}} - \sqrt{\kappa}\,\hat{p}-\sqrt{\gamma}\hat{m}.
\end{align}

Using this relation, the scattering(S) parameter, which characterizes the transmission of the system, can be obtained 

\begin{align}
S_{Trs} = \frac{\hat{p}_{\mathrm{out}}}{\hat{p}_{\mathrm{in}}}=
1
+
\frac{\kappa A_m + \gamma A_p + 2\sqrt{\kappa\gamma}\,G_{\mathrm{eff}}}
{A_pA_m-G_{\mathrm{eff}}^2} .
\end{align}

 \section{Population Dynamics}
 \label{app:Population dynamics}
\subsection{ Time Evolution of Mode Populations}
To understand how the system evolves from an initial state toward its steady
state, we analyze the time-dependent behavior of the photon and magnon modes under coherent driving.
The population of photon and magnon modes at time t is defined as 

\begin{align*}
n_p(t) &= \langle \hat{p}^\dagger(t)\hat{p}(t) \rangle, \\
n_m(t) &= \langle \hat{m}^\dagger(t)\hat{m}(t) \rangle .
\end{align*}

\subsubsection*{Time evolution of photon population}

To analyze the dynamical behavior of the system, we consider the time-dependent photon population defined as
\begin{equation}
n_p(t) = \langle \hat{p}^\dagger(t)\hat{p}(t) \rangle.
\end{equation}

The time evolution of the photon population is obtained using the Heisenberg equation of motion.
\begin{equation}
\frac{d}{dt} n_p(t) = \frac{d}{dt} \langle \hat{p}^\dagger \hat{p} \rangle
= \left\langle \frac{d}{dt}(\hat{p}^\dagger \hat{p}) \right\rangle.
\end{equation}

Applying the product rule, we obtain
\begin{equation}
\frac{d}{dt} n_p(t) =
\langle \dot{\hat{p}}^\dagger \hat{p} \rangle
+ \langle \hat{p}^\dagger \dot{\hat{p}} \rangle.
\end{equation}

The quantum Langevin equation governs the dynamics of the photon operator,
\begin{equation}
\dot{\hat{p}} =
-\left(i\delta_p + \beta + \kappa\right)\hat{p}
-i G \hat{m}
+ \sqrt{\kappa}\, p_{\text{in}},
\end{equation}
where \( G = g - i\Gamma e^{i(\Theta+\Phi)} \) represents the complex coupling strength. The Hermitian conjugate equation is given by
\begin{equation}
\dot{\hat{p}}^\dagger =
\left(i\delta_p - \beta - \kappa\right)\hat{p}^\dagger
+i G \hat{m}^\dagger
+ \sqrt{\kappa}\, p_{\text{in}}.
\end{equation}

Substituting these expressions into the time derivative of the photon population, we evaluate each term explicitly. The first term becomes,
\begin{equation}
\langle \dot{\hat{p}}^\dagger \hat{p} \rangle =
\left(i\delta_p - \beta - \kappa\right)\langle \hat{p}^\dagger \hat{p} \rangle
+i G \langle \hat{m}^\dagger \hat{p} \rangle
+ \sqrt{\kappa}\, p_{\text{in}} \langle \hat{p} \rangle,
\end{equation}
while the second term yields
\begin{equation}
\langle \hat{p}^\dagger \dot{\hat{p}} \rangle =
-\left(i\delta_p + \beta + \kappa\right)\langle \hat{p}^\dagger \hat{p} \rangle
- iG \langle \hat{p}^\dagger \hat{m} \rangle
+ \sqrt{\kappa}\, p_{\text{in}} \langle \hat{p}^\dagger \rangle.
\end{equation}

Adding these two contributions, the detuning terms cancel, and we obtain
\begin{align}
\frac{d}{dt} n_p(t) &=
-2(\beta+\kappa)\langle \hat{p}^\dagger \hat{p} \rangle
- iG (\langle \hat{p}^\dagger \hat{m} \rangle
-  \langle \hat{m}^\dagger \hat{p} \rangle \nonumber) \\
&\quad + \sqrt{\kappa}\, p_{\text{in}} \left( \langle \hat{p}^\dagger \rangle + \langle \hat{p} \rangle \right).
\end{align}

Defining the photon population \( n_p = \langle \hat{p}^\dagger \hat{p} \rangle \) and the intermode coherence \( X = \langle \hat{p}^\dagger \hat{m} \rangle \), the above equation can be written in compact form as
\begin{align}
\frac{d}{dt} n_p(t)
={}&
-2(\beta+\kappa)\, n_p(t)
-iG\left(X-X^\dagger\right)
\nonumber\\
&+
\sqrt{\kappa}\,p_{\mathrm{in}}
\left(
\langle\hat{p}^\dagger\rangle
+\langle\hat{p}\rangle
\right).
\end{align}

This equation shows that the photon population is not only governed by dissipation and external driving but is also intrinsically coupled to the magnon mode through the intermode coherence. Consequently, a complete description of the system requires simultaneous consideration of both population and coherence dynamics.
\subsubsection*{Time evolution of magnon population}

The time-dependent population of the magnon mode is defined as
\begin{equation}
n_m(t) = \left\langle \hat{m}^\dagger(t)\hat{m}(t) \right\rangle.
\end{equation}

The time evolution of the magnon population is obtained from the Heisenberg equation of motion,
\begin{equation}
\frac{d}{dt} n_m(t) = \frac{d}{dt} \langle \hat{m}^\dagger \hat{m} \rangle
= \left\langle \frac{d}{dt}(\hat{m}^\dagger \hat{m}) \right\rangle.
\end{equation}

Applying the product rule, we obtain
\begin{equation}
\frac{d}{dt} n_m(t) =
\langle \dot{\hat{m}}^\dagger \hat{m} \rangle
+ \langle \hat{m}^\dagger \dot{\hat{m}} \rangle.
\end{equation}

The quantum Langevin equation governs the dynamics of the magnon operator,
\begin{equation}
\dot{\hat{m}} =
-\left(i\delta_m + \alpha + \gamma\right)\hat{m}
- iG \hat{p}
+ \sqrt{\gamma}\, p_{\text{in}},
\end{equation}
while its Hermitian conjugate is given by
\begin{equation}
\dot{\hat{m}}^\dagger =
\left(i\delta_m - \alpha - \gamma\right)\hat{m}^\dagger
+i G \hat{p}^\dagger
+ \sqrt{\gamma}\, p_{\text{in}}.
\end{equation}

Substituting these expressions into the time derivative of the magnon population, we evaluate the two terms separately. The first term becomes
\begin{equation}
\langle \dot{\hat{m}}^\dagger \hat{m} \rangle =
\left(i\delta_m - \alpha - \gamma\right)\langle \hat{m}^\dagger \hat{m} \rangle
+i G \langle \hat{p}^\dagger \hat{m} \rangle
+ \sqrt{\gamma}\, p_{\text{in}} \langle \hat{m} \rangle,
\end{equation}
while the second term yields
\begin{equation}
\langle \hat{m}^\dagger \dot{\hat{m}} \rangle =
-\left(i\delta_m + \alpha + \gamma\right)\langle \hat{m}^\dagger \hat{m} \rangle
-i G \langle \hat{m}^\dagger \hat{p} \rangle
+ \sqrt{\gamma}\, p_{\text{in}} \langle \hat{m} \rangle,
\end{equation}

Adding these two contributions, the detuning terms cancel, yeilding
\begin{align}
\frac{d}{dt} n_m(t) &=
-2(\alpha+\gamma)\langle \hat{m}^\dagger \hat{m} \rangle
+ iG (\langle \hat{p}^\dagger \hat{m} \rangle
-  \langle \hat{m}^\dagger \hat{p} \rangle \nonumber) \\
&\quad + \sqrt{\gamma}\, p_{\text{in}} \left( \langle \hat{m}^\dagger \rangle + \langle \hat{m} \rangle \right).
\end{align}

Introducing the magnon population \( n_m = \langle \hat{m}^\dagger \hat{m} \rangle \) and the intermode coherence \( X = \langle \hat{p}^\dagger \hat{m} \rangle \), the above equation can be written in compact form as
\begin{equation}
\begin{aligned}
\frac{d}{dt} n_m(t)
={}&
-2(\alpha+\gamma)\,n_m(t)
+iG\left(X-X^\dagger\right)
\\
&+
\sqrt{\gamma}\,p_{\mathrm{in}}
\left(
\langle\hat{m}^\dagger\rangle
+\langle\hat{m}\rangle
\right).
\end{aligned}
\end{equation}

This result shows that the magnon population dynamics are also coupled to the photon mode through the intermode coherence, reflecting the hybrid nature of the coupled system.
\subsubsection*{Intermode coherence dynamics}

To obtain a closed set of equations, we introduce the intermode coherence defined as
\begin{equation}
X(t) = \langle \hat{p}^\dagger(t)\hat{m}(t) \rangle,
\quad
X^*(t) = \langle \hat{m}^\dagger(t)\hat{p}(t) \rangle.
\end{equation}

The time evolution of the coherence \( X(t) \) is obtained using the Heisenberg equation of motion,
\begin{equation}
\frac{d}{dt} X(t) = \frac{d}{dt} \langle \hat{p}^\dagger \hat{m} \rangle
= \left\langle \frac{d}{dt}(\hat{p}^\dagger \hat{m}) \right\rangle.
\end{equation}

Applying the product rule, we obtain
\begin{equation}
\frac{d}{dt} X(t) =
\langle \dot{\hat{p}}^\dagger \hat{m} \rangle
+ \langle \hat{p}^\dagger \dot{\hat{m}} \rangle.
\end{equation}

Substituting the quantum Langevin equations for the operators,
\begin{align}
\dot{\hat{p}}^\dagger &=
\left(i\delta_p - \beta - \kappa\right)\hat{p}^\dagger
+ iG \hat{m}^\dagger
+ \sqrt{\kappa}\, p_{\text{in}}, \\
\dot{\hat{m}} &=
-\left(i\delta_m + \alpha + \gamma\right)\hat{m}
- iG \hat{p}
+ \sqrt{\gamma}\, p_{\text{in}},
\end{align}
We evaluate each contribution.

The first term becomes
\begin{equation}
\langle \dot{\hat{p}}^\dagger \hat{m} \rangle =
\left(i\delta_p - \beta - \kappa\right)\langle \hat{p}^\dagger \hat{m} \rangle
+ iG \langle \hat{m}^\dagger \hat{m} \rangle
+ \sqrt{\kappa}\, p_{\text{in}} \langle \hat{m} \rangle,
\end{equation}
and the second term yields
\begin{equation}
\langle \hat{p}^\dagger \dot{\hat{m}} \rangle =
-\left(i\delta_m + \alpha + \gamma\right)\langle \hat{p}^\dagger \hat{m} \rangle
- iG \langle \hat{p}^\dagger \hat{p} \rangle
+ \sqrt{\gamma}\, p_{\text{in}} \langle \hat{p}^\dagger \rangle.
\end{equation}

Adding these contributions, we obtain
\begin{align}
\frac{d}{dt} X(t) &=
\left[-i(\delta_m - \delta_p) - (\beta+\kappa+\alpha+\gamma)\right] X(t) \nonumber \\
&\quad - iG\, \langle \hat{p}^\dagger \hat{p} \rangle
+ iG\, \langle \hat{m}^\dagger \hat{m} \rangle \nonumber \\
&\quad + \sqrt{\kappa}\, p_{\text{in}} \langle \hat{m} \rangle 
+ \sqrt{\gamma}\, p_{\text{in}} \langle \hat{p}^\dagger \rangle.
\end{align}

Introducing the populations $n_p = \langle \hat{p}^\dagger \hat{p} \rangle$ and $n_m = \langle \hat{m}^\dagger \hat{m} \rangle$, the equations can be written as
\begin{align}
\frac{d}{dt} X(t) = {} &
\left[-i\Delta - \kappa_{T}\right] X(t)
- iG(n_p - n_m) \nonumber \\
& + \sqrt{\kappa}\, p_{\text{in}} \langle \hat{m} \rangle
+ \sqrt{\gamma}\, p_{\text{in}} \langle \hat{p}^\dagger \rangle,
\\[1ex]
\frac{d}{dt} X^\dagger(t) = {} &
\left[i\Delta - \kappa_{T}\right] X^\dagger(t)
+ iG(n_p - n_m) \nonumber \\
& + \sqrt{\kappa}\, p_{\text{in}} \langle \hat{m}^\dagger \rangle
+ \sqrt{\gamma}\, p_{\text{in}} \langle \hat{p} \rangle,
\end{align}

where $\Delta = \delta_m - \delta_p$ is the detuning difference and $\kappa_T= \beta+\kappa+\alpha+\gamma $ are the total effective damping of the system. These equations demonstrate that the intermode coherence is directly coupled to both photon and magnon populations, thereby completing the closed set of dynamical equations for the system.

%

\end{document}